\documentclass[12pt]{article}

\usepackage{amsmath, amssymb, mathtools}
\usepackage{type1cm}
\usepackage{mathrsfs}
\usepackage{physics}
\usepackage[T1]{fontenc}
\usepackage{fancyhdr}
\usepackage{appendix}
\usepackage{titlesec}
\usepackage{cite}
\usepackage{color}
\usepackage[top=25truemm,bottom=25truemm,left=20truemm,right=20truemm]{geometry}
\usepackage[dvipdfmx]{graphicx}
\usepackage{bm}

\baselineskip=\normalbaselineskip

\newcommand\eq[1]{\begin{equation}#1\end{equation}}
\newcommand\al[1]{\begin{align}#1\end{align}}

\newcommand\als[1]{\begin{align}\begin{split}#1\end{split}\end{align}}
\newcommand{\im}{\mathrm{i}}
\newcommand{\dif}{\mathrm{d}}
\newcommand{\e}{\mathrm{e}}

\makeatletter
 
 \@addtoreset{equation}{section}
\makeatother

\begin{document}
\setcounter{footnote}{0}
\setcounter{tocdepth}{3}
\bigskip
\def\thefootnote{\arabic{footnote}}

\begin{titlepage}
\renewcommand{\thefootnote}{\fnsymbol{footnote}}
\begin{normalsize}
\begin{flushright}
\begin{tabular}{l}
UTHEP-774
\end{tabular}
\end{flushright}
  \end{normalsize}

~~\\

\vspace*{0cm}
    \begin{Large}
       \begin{center}
         {Vector bundles on fuzzy K\"{a}hler manifolds}
       \end{center}
    \end{Large}

\vspace{0.7cm}

\begin{center}
Hiroyuki A\textsc{dachi}$^{1)}$\footnote[1]
            {
e-mail address : 
adachi@het.ph.tsukuba.ac.jp},
Goro I\textsc{shiki}$^{1),2)}$\footnote[2]
            {
e-mail address : 
ishiki@het.ph.tsukuba.ac.jp}
and
Satoshi K\textsc{anno}$^{1)}$\footnote[3]
            {
e-mail address : 
kanno@het.ph.tsukuba.ac.jp}

\vspace{0.7cm}

     $^{ 1)}$ {\it Graduate School of Science and Technology, University of Tsukuba, }\\
               {\it Tsukuba, Ibaraki 305-8571, Japan}\\

     $^{ 2)}$ {\it Tomonaga Center for the History of the Universe, University of Tsukuba, }\\
               {\it Tsukuba, Ibaraki 305-8571, Japan}\\
               \end{center}

\vspace{0.5cm}

\begin{abstract}
\noindent
We propose a matrix regularization of vector bundles over a general closed K\"ahler manifold.
This matrix regularization is given as a natural generalization of the Berezin-Toeplitz quantization and gives a map from sections of a vector bundle to matrices.
We examine the asymptotic behaviors of the map in the large-$N$ limit.
For vector bundles with algebraic structure, we derive a beautiful correspondence of the algebra of sections and the algebra of corresponding matrices in the large-$N$ limit.
We give two explicit examples for monopole bundles over a complex projective space $CP^n$ and a torus $T^{2n}$.

\end{abstract}

\end{titlepage}

\tableofcontents

\section{Introduction}

The notion of noncommutative geometry appears in various studies of superstring theory and M-theory
\cite{Seiberg:1999vs,Banks:1996vh,Ishibashi:1996xs} 
and it suggests that the noncommutative geometry might be suitable to describe the space-time in Planck scale rather than a smooth manifold.
In noncommutative geometry, we consider the space-time coordinates as a set of noncommutative operators on some Hilbert space.
A particular family of noncommutative geometry is called fuzzy geometry, which is the case when the Hilbert space is finite-dimensional and the space-time coordinates are finite dimensional square matrices.
This fuzzy geometry plays an important role in matrix models of superstring theory and M-theory.

In order to describe physics on such fuzzy geometry, it is needed to formulate various fields on this geometry.
For example, to describe the low energy effective theories of D-branes, we need the fuzzy description of the field theories in the matrix models.
For this purpose, it is important to find a description of a fuzzy version of vector bundles, since ordinary fields are described as sections of some vector bundles.
The motivation of this paper is to generalize a matrix regularization \cite{Hoppe}, which is a map from functions on a smooth manifold to corresponding matrices on a fuzzy geometry.
More specifically, we establish matrix regularization of vector bundles over a connected closed K\"{a}hler manifold.

Conventionally, the matrix regularization of functions on a closed symplectic manifold is described in the following manner.
Let us consider a closed $2n$-dimensional symplectic manifold $(M,\omega)$.
From the symplectic structure $\omega$, one can define a volume form $\mu := \omega^{\wedge n}/n!$ and a Poisson bracket
\eq{ \{f, g \} := W^{\mu\nu} \partial_{\mu}f \partial_{\nu} g ,}
where $f,g$ are smooth functions and $W^{\mu\nu}$ is the Poisson tensor defined by $\omega_{\mu\nu}W^{\mu\rho} = \delta_{\nu}^{\rho}$.
Let $\{N_p\}$ be a sequence of strictly increasing integers satisfying $N_p \to \infty$ as $p\to \infty$.
The matrix regularization is defined as a sequence of linear maps
$T_p:C^{\infty}(M) \to M_{N_p}(\mathbb{C})$ which satisfies \cite{Arnlind:2010ac}
\als{
&\lim_{p\to \infty} |T_p(f)T_p(g)-T_p(fg) | = 0,\\
&\lim_{p\to \infty} |\im \hbar_p^{-1}[T_p(f), T_p(g)]- T_p(\{f,g \})| = 0,\\
&\lim_{p\to \infty} (2\pi \hbar_p)^n \mathrm{Tr} \, T_p(f) = \int_M \mu f.
\label{MatReg}
}
Here, $\hbar_p = (kp)^{-1}$ for some constant $k$ and $|\cdot|$ is a matrix norm.
These conditions can be seen as an analogue of 
the canonical quantization of classical mechanics where the phase space is $T^* \mathbb{R}^{n} \simeq \mathbb{R}^{2n}$.
These relations are essential in deriving the action of the matrix model from 
the worldvolume action of a membrane \cite{Hoppe}.

For a symplectic manifold $M$, it is known that there indeed exists a map $T_p$ satisfying (\ref{MatReg}).
A systematic and beautiful construction of such a map is given by the Berezin-Toeplitz quantization \cite{Bordemann:1993zv, Ma-Marinescu}.
In this quantization, we first consider a suitable Dirac operator with $N_p$ zero modes \cite{Ma-Marinescu}.
Then, one defines $T_p(f)$ by $T_p(f) = \Pi f \Pi$,
where $\Pi$ is the projection operator onto the Dirac zero modes.
The map $T_p$, sometimes referred to as the Toeplitz operator, indeed satisfies all the properties of (\ref{MatReg}).

The Toeplitz operators for more general fields than functions were proposed in \cite{Hawkins:1997gj,Hawkins:1998nj,Hawkins:2005,Adachi:2020asg,Nair:2020xzn}.
In more recent studies \cite{Adachi:2021ljw, Adachi:2021aux}, it is shown that the Toeplitz operator of general fields on a closed Riemann surface enjoys beautiful properties, which are a natural generalization of (\ref{MatReg}).

In this paper, we investigate the Berezin-Toeplitz quantization of vector bundles over a general closed K\"ahler manifold.
We show that the asymptotic properties of the Toeplitz operator given in \cite{Adachi:2021ljw, Adachi:2021aux} also exist in higher dimensional manifolds. 
We derive a large-$p$ asymptotic expansion of the product $T_p(\varphi) T_p(\chi)$ for arbitrary sections of vector bundles (general fields) $\varphi,\chi$, up to the second order in $1/p$.
From this asymptotic expansion, we obtain important relations of the Toeplitz operator including generalization of (\ref{MatReg}).
We also give explicit examples of monopole bundles over a fuzzy $CP^n$ \cite{Carow-Watamura:2004mug,Dolan:2006tx} and fuzzy $T^{2n}$ \cite{Adachi:2020asg}, where the Dirac operator zero modes have relatively simple representations
\footnote{See \cite{Honda:2020ckp} for the analysis of Dirac operator zero modes of Riemann surfaces with higher genera, where the zero modes have more complex representations than those of $CP^1=S^2$ and $T^2$.}.

This paper is organized as follows. 
In section 2, we propose the Berezin-Toeplitz quantization for general vector bundles and 
derive the asymptotic expansion.
In section 3 and 4, we consider the Berezin-Toeplitz quantization of monopole bundles over $CP^n$ and $T^{2n}$, respectively.
In section 5, we give a summary and a discussion.

\section{Berezin-Toeplitz quantization}

In this section, we consider the Berezin-Toeplitz quantization for vector bundles and 
derive an asymptotic expansion of the quantization map.
In subsection 2.1, we define the Toeplitz operator for vector bundles.
In subsection 2.2, we derive the asymptotic behaviors of the Toeplitz operators.
In subsection 2.3, we show the relation between the trace of the Toeplitz operator and the integral of the corresponding field in the large-$N$ limit.
In subsection 2.4, we construct the matrix Laplacian.

\subsection{Berezin-Toeplitz quantization for vector bundles}

We consider a closed connected $2n$-dimensional K\"{a}hler manifold $M$ with a K\"{a}hler structure $(g,J,\omega)$, where $g$ is a Riemannian metric, $J$ is a complex structure and $\omega$ is a symplectic form satisfying the compatibility condition,
\eq{
\label{Kahler}
\omega(\cdot,\cdot)=g(J\cdot,\cdot).
}
The K\"{a}hler potential $K$ is a function defined by the local relation $\omega = \im \partial \bar{\partial}K$ where $\partial, \bar{\partial}$ are Dolbeault differential operators.
A natural volume form is defined by $\mu := \omega^{\wedge n}/n!$.
In terms of the local real coordinates $\{x^{\mu}\}_{\mu =1}^{2n}$, we have $\mu = \sqrt{g}\, \dif x^1 \wedge \dif x^2 \wedge \cdots \wedge \dif x^{2n}$.
To define the quantization map, we will introduce three Hermitian vector bundles $L$ , $S_c$ and $E$.
$L$ is a prequantum line bundle, $S_c$ is a spin-$c$ bundle and $E$ is the target bundle which we want to quantize.
$L$ can be defined for a quantizable manifold, which we will discuss below, and $S_c$ is known to exist for any K\"ahler manifold.
For any vector bundle $F$, we will denote the connection and the curvature of $F$ by $\nabla^F = \dif + A^F$ and $R^F := (\nabla^F)^2 = \dif A^F + A^F \wedge A^F$, respectively, where $A^F$ is the connection one-form of $F$.

A prequantum line bundle $L$ is a complex line bundle with a connection $\nabla^L$ such that its curvature (field strength) $R^L$ is proportional to the symplectic form:
\eq{
\label{k}
    R^L = -\im k \omega.
}
Here, the constant factor $k$ is chosen such that $\frac{\im}{2\pi}\int_{\Sigma} R^L \in \mathbb{Z}$, where $\Sigma \subseteq M$ is any two-cycle of $M$.
This condition is equivalent to the condition that the symplectic form $\frac{k}{2\pi}\omega$ is in the second integer cohomology $H^2(M,\mathbb{Z})$.
Manifolds which allow the existence of this prequantum line bundle are called quantizable manifold.
For a two dimensional manifold $M=\Sigma$, we can take $k=2\pi/\int_M \omega$.
The connection one-form $A^L$ is defined by the local expression of the connection $\nabla^L = \dif + A^L$.
Using the K\"{a}hler potential $K$, one can choose a connection one-form by
\eq{
\label{A^L from K}
A^L = -\frac{k}{2} (\partial - \bar{\partial})K.}
Let $\Gamma(\cdot)$ be a set of all the smooth sections of the vector bundle.
Then, an element of $\Gamma(L)$ is a smooth complex scalar fields coupling to a $\mathrm{U}(1)$ background gauge field $A^{L}$.
For the two-dimensional case, the curvature is proportional to the volume form, which means that sections of $L$ are complex scalar fields coupling to uniform magnetic flux.

Next, we consider the spin-$c$ structure (see \cite{BTQ textbook,spin geometry} for more rigorous mathematical treatment).
The canonical spin-$c$ bundle is defined by $S_c := \bigoplus_{p=0}^n \Lambda^{0,p}(T^*M)$, i.e. its fiber is a sum of $(0,p)$-forms.
This bundle is formally equal to $S \otimes L_c^{1/2}$, where $S$ is the canonical spin bundle and $L_c$ is the determinant line bundle of holomorphic tangent bundle $L_c:= \det T^{(1,0)}M$.
In the case of non-spin manifold, $S$ and the square root bundle $L_c^{1/2}$ themselves are not well-defined and only the tensor product $S_c = S \otimes L_c^{1/2}$ is well-defined
\footnote{
Precisely speaking, though both $S$ and $L_c^{1/2}$ can be locally defined, the cocycle conditions of the transition functions are not satisfied for non-spin manifolds.
However, the violations of the cocycle conditions cancel out for the formal tensor product $S_c=S\otimes L_c^{1/2}$, so that $S_c$ is globally well-defined.
$CP^{2m} \ (m\in \mathbb{N})$ is an example of non-spin manifold with the spin-$c$ structure.}.
A connection of $S_c$ is locally given by $\nabla^{S_c} = \dif + A^S + \frac{1}{2}A^{L_c}$.
The connection one-form of the canonical spin bundle $S$ is defined by
\eq{ A^S = \frac{1}{4} \gamma_{(2n)}^a  \gamma_{(2n)}^b \Omega_{ab}}
where $\{\gamma_{(2n)}^a \}_{a=1}^{2n}$ is a set of gamma matrices satisfying the Clifford algebra $\{ \gamma_{(2n)}^a  ,\gamma_{(2n)}^b \} = 2 \delta_{ab} I_{2^n}$ discussed in Appendix \ref{Clifford} and $\Omega_{ab} = \Omega_{ab\mu} \dif x^{\mu}$ is the spin connection one-form
\eq{\Omega_{ab\mu} = e_a{}^{\nu} g_{\nu \rho}  (\partial_{\mu} e_{b}{}^{\rho} + \Gamma^{\rho}_{\mu \sigma} e_b{}^{\sigma}) .}
Here, $\{e_a\}_{a=1}^{2n}$ is a set of the local orthonormal frame fields (vielbeins) satisfying $g(e_a,e_b) = \delta_{ab}$.
The connection one-form of $L_c$ is given by $A^{L_c} = - \sum_{m=1}^n \Omega_{m \bar{m}}$, where $m$ and $\bar{m}$ are indices of complexified orthonormal frame vector fields introduced in (\ref{complex orthonormal}).
We can interpret the sections of $S_c$ as complex spinor fields coupling to $\frac{1}{2}A^{L_c}$.

Now, we consider the target bundle $E$.
We assume that $E$ is a finite-rank Hermitian vector bundle. 
We express $E$ as a homomorphism bundle (Hom-bundle) $\mathrm{Hom}(E_2,E_1)$, where $E_i \,(i=1,2)$ are some Hermitian vector bundles.
Here, $\mathrm{Hom}(E_2,E_1)$ is a vector bundle whose fiber at a point $x\in M$ is a vector space of linear maps from the fiber of $E_2$ at $x$ to the fiber of $E_1$ at $x$.
Note that any vector bundles can always be written as the Hom-bundle.
The reason why we introduce the Hom-bundle is to introduce an algebraic structure which we will quantize.
Namely, there is a natural product structure between $\Gamma(\mathrm{Hom}(E_2,E_1)) \times \Gamma(\mathrm{Hom}(E_3,E_2)) \to \Gamma(\mathrm{Hom}(E_3,E_1))$, following from the pointwise composition of the linear maps.
This product is mapped to the product of matrices in the quantization we discuss below.

The description using the Hom-bundle is applicable to most fields appearing in physics.
For example, let $\tilde{L}$ be a complex line bundle with connection one-form $A^{\tilde{L}}$.
Then, $\tilde{L}^{\otimes q}$ can be written as $\mathrm{Hom}(\tilde{L}^{\otimes r},\tilde{L}^{\otimes q+ r})$ for any integers $q,r$.
This means that a section of $\tilde{L}^{\otimes q}$, which is a complex scalar field coupling to $A^{\tilde{L}}$ with charge $q$, can also be regarded as a linear map from fields with charge $r$ to those with charge $q+r$.
Another example is that adjoint matter fields are regarded as linear maps from fundamental matter fields to themselves.
Finally, tensor fields can also be viewed as linear maps between tensor fields with various ranks.
For instance, a section of $\mathrm{Hom}((TM)^{\otimes r},(TM)^{\otimes q})$ is a tensor field of $(q,r)$ type:
\eq{ (\varphi_1)^{\mu_1 \mu_2 \cdots \mu_q} = (\varphi)^{\mu_1 \mu_2 \cdots \mu_q}{}_{\nu_1 \nu_2 \cdots \nu_r} (\varphi_2)^{\nu_1 \nu_2 \cdots \nu_r},}
which corresponds to $(TM)^{\otimes q} \otimes (T^*M)^{\otimes r} \simeq \mathrm{Hom}((TM)^{\otimes r},(TM)^{\otimes q})$.

As we have discussed, $\Gamma(\mathrm{Hom}(E_2,E_1))$ can be thought of a linear map $\Gamma(E_2) \to \Gamma(E_1)$.
We can extend this linear structure to a map $\Gamma(S_c \otimes L^{\otimes p} \otimes E_2) \to \Gamma(S_c \otimes L^{\otimes p} \otimes E_1)$ by just acting as an identity on fibers of the auxiliary bundle $S_c \otimes L^{\otimes p}$ at each point $x \in M$.
Here, $p$ is an integer.
Note that $\Gamma(S_c\otimes L^{\otimes p} \otimes E_i)$ are infinite dimensional vector spaces.
If we can restrict this linear map to be a map between finite-dimensional subspaces, such a map can be regarded as a finite dimensional matrix.
This is the main idea of the Berezin-Toeplitz quantization.
In order to realize such a scenario, let us consider Dirac operators on $\Gamma(S_c \otimes L^{\otimes p} \otimes E_i)$ by
\eq{
    \label{twisted Dirac}
    D_i = \im \gamma_{(2n)}^{a} \nabla^{S_c\otimes L^{\otimes p} \otimes E_i}_{e_a} = \im e_a{}^{\mu}\gamma_{(2n)}^{a} \left( \partial_{\mu} + \frac{1}{4} \Omega_{ab\mu} \gamma_{(2n)}^a \gamma_{(2n)}^b - \frac{1}{2} \sum_{m=1}^n \Omega_{m \bar{m} \mu} +p A^L_{\mu} + A^E_{\mu} \right) .
}
We equip an inner product on $\Gamma(S_c \otimes L^{\otimes p} \otimes E_i)$ by
\eq{
    \label{inner product}
    (\psi',\psi) := \int_M \mu \, (\psi')^{\dagger} \cdot \psi \quad (\psi,\psi' \in \Gamma(S_c \otimes L^{\otimes p} \otimes E_i))
}
where $(\psi')^{\dagger} \cdot \psi$ is a Hermitian inner product of a fiber $S_c \otimes L^{\otimes p} \otimes E_i$ at point $x \in M$, which is defined by a combination of Hermitian metrics of $S_c, L$ and $E_i$.
In the physicist language, $\dagger$ and $\cdot$ simply mean the Hermitian conjugation and the contractions of indices, respectively.
The norm is defined by 
$|\psi| = \sqrt{(\psi,\psi)}$.
The space of normalizable zero modes $\mathrm{Ker}\,D_i$ is finite dimensional.
With the particular choice of the gamma matrices in Appendix \ref{Clifford}, one can compute its dimension $N_i := \dim \mathrm{Ker} D_i$ for sufficiently large $p$ using the Atiyah-Singer index theorem and the vanishing theorem as shown in \ref{Vanishing}.
Here, $p$ controls the dimension $N_i$ where $N_i$ plays the role of matrix size of the matrix regularization map.
Now, let $\Pi_i$ be a projection from $\Gamma(S _c\otimes L^{\otimes p} \otimes E_i)$ to $\mathrm{Ker} D_i$.
We define the Berezin-Toeplitz quantization of $\Gamma(\mathrm{Hom}(E_2,E_1))$ by
\eq{
    T_p^{(E_1,E_2)} (\varphi) = \Pi_1 \varphi \Pi_2 \quad (\varphi \in \Gamma(\mathrm{Hom}(E_2,E_1))).
    \label{Toeplitz op}
}
Here, $T_p^{(E_1,E_2)} (\varphi)$ is a map from $\mathrm{Ker} D_2$ to $\mathrm{Ker} D_1$ and therefore it can be represented as an $N_1 \times N_2$ matrix.
As we will see below, the Toeplitz operator (\ref{Toeplitz op}) enjoys a nice asymptotic behavior,
which gives a generalization of (\ref{MatReg}).

\subsection{Asymptotic expansion of Toeplitz operators}
We can also consider another bundle $\mathrm{Hom}(E_3,E_2)$ and define a Toeplitz operator $T_p^{(E_2,E_3)} (\chi) = \Pi_2 \chi \Pi_3$ for $\chi \in \Gamma(\mathrm{Hom}(E_3,E_2))$.
Then, we can consider a product $T_p^{(E_1,E_2)} (\varphi) T_p^{(E_2,E_3)} (\chi)$.
As shown in Appendix \ref{asymptotic expansion}, the Toeplitz operator (\ref{Toeplitz op}) has
the following asymptotic expansion in $\hbar_p = (kp)^{-1}$:
\eq{
     T_p^{(E_1,E_2)} (\varphi) T_p^{(E_2,E_3)} (\chi) = \sum_{i=0}^{\infty} \hbar_{p}^i T_p^{(E_1,E_3)} (C_i(\varphi,\chi)),
    \label{asym exp}
}
where the symbols $C_i$ on the right-hand side are maps from $\Gamma(\mathrm{Hom}(E_2,E_1)) \times \Gamma(\mathrm{Hom}(E_3,E_2))$ to $\Gamma(\mathrm{Hom}(E_3,E_1))$.
We find that the first three $C_i$'s are explicitly given by 
\als{
C_0(\varphi,\chi) &= \varphi \chi,\\
C_1(\varphi,\chi) &= -\frac{1}{2} G^{\alpha\beta}(\nabla_{\alpha}\varphi)(\nabla_{\beta}\chi),\\
C_2(\varphi,\chi) &= \frac{1}{8} G^{\alpha\beta}G^{\gamma\delta} [(\nabla_{\alpha} \varphi) ( \im R_{\beta \gamma \mu \nu} W^{\mu\nu} - 2 R^{E_2}_{\beta \gamma}) (\nabla_{\delta} \chi) + (\nabla_{\alpha} \nabla_{\gamma} \varphi) (\nabla_{\beta} \nabla_{\delta} \chi)].
\label{asymptotic exp}
}
Here, we introduced a tensor $G^{\alpha\beta} := g^{\alpha\beta} + \im W^{\alpha\beta}$, where $g^{\alpha\beta}$ is the inverse of the metric tensor and $W^{\alpha\beta}$ is a Poisson tensor defined by $\omega_{\mu\nu}W^{\mu\rho} = \delta_{\nu}^{\rho}$.
In (\ref{asymptotic exp}), $R_{\alpha \beta \gamma \delta}$ is the Riemann curvature tensor for the metric $g$ and $R^{E_2}_{\alpha \beta} := R^{E_2}(\partial_{\alpha}, \partial_{\beta})$ is a component of the curvature of $E_2$.
The operator $\nabla_{\alpha}$ is the covariant derivative on each field.
For example, it acts on $\varphi \in \Gamma(\mathrm{Hom}(E_2,E_1))$ as $\nabla_{\alpha}\varphi = \partial_{\alpha} \varphi + A^{E_1}_{\alpha} \varphi - \varphi A^{E_2}_{\alpha}$, where $A^{E_i}$ is a connection one-form of $E_i$.
In Appendix \ref{Consistency check of the asymptotic expansion}, we checked that (\ref{asymptotic exp}) is consistent with the associativity of the operator product.

We leave the proof of (\ref{asym exp}) to Appendix \ref{asymptotic expansion} and discuss here some important corollaries of (\ref{asymptotic exp}).
From (\ref{asymptotic exp}), it is easy to show the following relation:
\eq{
\label{generalized matrix regularization 1}
\lim_{p\to\infty}\left|T_p^{(E_1,E_2)}(\varphi )T_p^{(E_2,E_3)}(\chi) - T_p^{(E_1,E_3)}(\varphi\chi)\right|=0.
}
Moreover, let us consider a function $f \in C^{\infty}(M)$ and identity operator $\mathbf{1}_{E_i} \in \Gamma(\mathrm{End}(E_{i}))$. 
Then, we can consider the following commutator-like operation:
\eq{
\label{generalized commutator}
[T(f \mathbf{1}) , T_p^{(E_1,E_2)}(\varphi)] := T_p^{(E_1,E_1)}(f \mathbf{1}_{E_1})  T_p^{(E_1,E_2)}(\varphi) - T_p^{(E_1,E_2)}(\varphi) T_p^{(E_2,E_2)}(f \mathbf{1}_{E_2}).
}
Using the asymptotic expansion to (\ref{generalized commutator}), one finds
\eq{
\label{generalized matrix regularization 2}
\lim_{p\to\infty} \left|\im \hbar_{p}^{-1}[T(f \mathbf{1}) , T_p^{(E_1,E_2)}(\varphi)] -T_p^{(E_1,E_2)}(\{f,\varphi\})\right|=0,
}
where the generalized (covariantized) Poisson bracket $\{\;\; ,\;\; \}$ is defined by 
\eq{
\label{generalized Poisson bracket}
\{f ,\varphi\} := W^{\alpha\beta}(\partial_{\alpha}f)  (\nabla_{\beta}\varphi).
}
From this correspondence, one can express the covariant derivative on $\varphi$ by this commutator-like operation in matrix models.

For the trivial line bundle $E_i = M\times \mathbb{C}$, i.e. for ordinary complex valued functions and for simple pointwise products, the relations (\ref{generalized matrix regularization 1}) and (\ref{generalized matrix regularization 2})
reduce to the first two in (\ref{MatReg}).

\subsection{Trace of the Toeplitz operator}

Let us consider the case for an endomorphism bundle $\mathrm{End}(E_1) = \mathrm{Hom}(E_1,E_1)$.
Then, we can consider the Toeplitz operator of $\varphi \in \Gamma(\mathrm{End}(E_1))$ given by
\eq{
T_p^{(E_1,E_1)}(\varphi) = \Pi_1 \varphi \Pi_1.
}
In this case, we can define a trace of the Toeplitz operator.
As shown in Appendix~\ref{trace}, we obtained the following property
\eq{
\label{trace property}
\lim_{p \to \infty} (2\pi\hbar_p)^n \Tr T_p^{(E_1,E_1)}(\varphi) = \int_M  \mu \, \mathrm{tr}_{E_1} \varphi.
}
Here, $\mathrm{tr}_{E_1}$ is a trace in terms of vector space of the fiber of $E_1$.
This result is a generalization of the third equation in (\ref{MatReg}).

\subsection{Bochner Laplacian and its matrix regularization}

Let $E$ be a Hermitian vector bundle over $M$ and let $\nabla^E : \Gamma(E) \to \Gamma(E \otimes T^*M)$ be a Hermitian connection of $E$.
Let us also consider the adjoint of the connection $(\nabla^E)^* :\Gamma(E \otimes T^*M) \to  \Gamma(E)$.
Then, the Bochner Laplacian $\Delta^E$ is defined by
\eq{
\Delta^E \varphi := (\nabla^E)^* \nabla^E \varphi.
\label{def of Laplacian}
}
In terms of the local coordinate, we write
\eq{
\Delta^E \varphi = -g^{\mu\nu} \nabla_{\mu} \nabla_{\nu} \varphi,
}
where the first covariant derivative is simply equal to $\nabla_{\nu} \varphi = (\partial_{\nu} + A^E_{\nu})\varphi$ but the second covariant derivative acts on $\nabla_{\nu} \varphi$ as $\nabla_{\mu} \nabla_{\nu} \varphi = (\partial_{\mu} + A^E_{\mu}) \nabla_{\nu} \varphi - \Gamma^{\rho}_{\mu\nu}\nabla_{\rho} \varphi$.
If a sections of $E$ has an additional orthonormal index, the covariant derivative is assumed to be $\nabla_{\mu} \varphi_a = (\partial_{\mu} + A^E_{\mu})\varphi_a + \Omega_{ab \mu} \varphi_b$.
In this notation, we have $\nabla_{\mu} e_{a}^{\nu} =0$ and $\nabla_{\mu} \gamma^a_{(2n)} =0$.
Also, let us introduce $\nabla_a := \nabla_{e_a} = e_a{}^{\mu} \nabla_{\mu}$.
Then, we have useful identities $\Delta^E = - \nabla_a \nabla_a$ and $[\nabla_a,\nabla_b] \varphi = R^E(e_a,e_b) \varphi$
\footnote{There is also another expression $\Delta^E = - (\nabla^E_{e_a})^2 + \nabla^E_{\nabla^{TM}_{e_a} e_a}$ and $([\nabla^E_{e_a},\nabla^E_{e_b}] - \nabla^E_{[e_a,e_b]})\varphi = R^E(e_a,e_b) \varphi$, which we can find in mathematical literature.}.

In order to construct the matrix Laplacian, let us consider the following trick.
Let $\{X^A\}_{A=1,2,\cdots,d}$ be isometric embedding coordinate functions satisfying
\eq{
(\partial_{\mu} X^A)(\partial_{\nu} X^A) =g_{\mu\nu},
}
where the existence of such an embedding is ensured by Nash embedding theorem for sufficiently large $d$.
As shown in \ref{Laplace operator}, the Laplacian can be written by using 
the isometric embedding functions and covariant Poisson bracket:
\eq{
\label{laplacian embedding}
    \Delta^E \varphi = - \{X^A,\{X^A,\varphi \}\}.
}
This expression is given in terms of the generalized Poisson bracket, it is easy to find the corresponding matrix Laplacian.

From (\ref{laplacian embedding}), it is natural to define the matrix Laplacian $\hat{\Delta}$ by
\eq{
\hat{\Delta} T_p^{(E_1,E_2)}(\varphi) := \hbar_p^{-2}[T(X^A{\bf 1}),[T(X^A{\bf 1}), T_p^{(E_1,E_2)}(\varphi)]],
\label{fuzzy matrix laplacian}
}
for $\varphi \in \Gamma({\rm Hom}(E_2,E_1))$.
Here, $[\;\; ,\;\; ]$ is the generalized commutator defined in (\ref{generalized commutator}).
We can see that $\hat{\Delta}$ is a Hermitian operator which is positive semi-definite in terms of the Frobenius inner product.
In \cite{Adachi:2021ljw}, it is shown that the spectra of the Bochner and the matrix Laplacians 
agree in the large-$p$ limit
\footnote{This is explicitly shown for the case $\dim M =2$ \cite{Adachi:2021ljw} and the proof can be easily generalized in the case of general K\"ahler manifold we are considering in this paper.}.

\section{Fuzzy $CP^n$}

In this section, we consider a Berezin-Toeplitz quantization of monopole bundle over a complex projective space $CP^n$.
Other construction of such quantization map are given in \cite{Carow-Watamura:2004mug,Dolan:2006tx}.
In subsection 3.1, we define a complex projective space $CP^n$ and describe basic properties. 
In subsection 3.2, we explicitly construct a complete orthonormal basis of the kernel of the Dirac operator. 
In subsection 3.3, we calculate Toeplitz operators of embedding functions. 
In subsection 3.4 and 3.5, we discuss the continuum Laplacian and the matrix Laplacian, respectively, for a monopole bundle~\footnote{
The correspondence of matrices and (charged) fields was studied in \cite{Carow-Watamura:2004mug,Dolan:2006tx}, where they use the projective module construction and the Fock space construction.
In particular, the correspondence of Laplacians is extensively studied in \cite{Dolan:2006tx}.
In our formalism, the underlying mechanism of these correspondences is revealed based on the asymptotic expansion of the Toeplitz operators.
Furthermore, our formalism can be applied to any general K\"ahler manifolds and any vector bundles.}.

\subsection{Geometry of $CP^n$}
Firstly, let us define $CP^n$, which is a closed connected $2n$-dimensional K\"{a}hler manifold.
For $Z,Z' \in \mathbb{C}^{n+1}\setminus \{0\}$, we will define a equivalence relation $\sim$ by
\eq{Z\sim Z' \quad :\Leftrightarrow \quad {}^{\exists}  \, c \in \mathbb{C}\setminus \{0\}: \ Z=cZ'.}
Then, $CP^n$ is defined by
\eq{CP^n = (\mathbb{C}^{n+1} \setminus \{0\})/\sim.}
This space can be covered by a set of $n+1$ patches $\{U_{\alpha}\}_{\alpha=1}^{n+1}$ where $U_{\alpha} := \{ [Z] \in CP^n \mid Z^{\alpha}\neq 0\}$.
Here, $[Z]=[Z^1,Z^2,\cdots,Z^{n+1}]$ is a representative class with respect to the relation $\sim$ and is called homogeneous coordinates.
For a patch $U_{\alpha}$, one can define inhomogeneous coordinates $(z_{(\alpha)}^{1}, z_{(\alpha)}^{2} , \cdots, z_{(\alpha)}^{n})$ such that
\eq{
z_{(\alpha)}^{\mu} =
\begin{cases}
Z^{\mu}/Z^{\alpha} & (\mu = 1,2,\cdots,\alpha-1)\\
Z^{\mu+1}/Z^{\alpha} & (\mu = \alpha,\alpha+1,\cdots,n)
\end{cases}
.
}

In order to define a K\"ahler structure of $CP^n$, let us consider a local function $K_{\alpha}$ on a patch $U_{\alpha}$ as
\eq{K_{\alpha}(p) := \log (1+\sum_{\mu=1}^n |z_{(\alpha)}^{\mu}(p)|^2) = \log(\sum_{\mu=1}^{n+1} |Z^{\mu}/Z^{\alpha}|^2) .}
For $x \in U_{\alpha} \cap U_{\beta}$, we have
\eq{K_{\alpha}(x) = K_{\beta}(x) + \log(Z^{\beta}/Z^{\alpha})+ \log (\overline{Z^{\beta}/Z^{\alpha}}).}
By acting the Dolbeault differentials $\partial,\bar{\partial}$, we have $\partial \bar{\partial} K_{\alpha} = \partial \bar{\partial} K_{\beta}$.
Thus, we can define a closed two-form $\omega$ locally written as
\eq{\label{K on CP^n}
\omega = \im \partial \bar{\partial} K.}
From now on, we will omit the subscripts of the patch.
By using the local complex coordinates $z^{\mu}$, $\omega$ is written as
\eq{\omega = \im \frac{(1+ |z|^2) \delta_{\mu\nu} - \bar{z}^{\mu} z^{\nu}}{(1+ |z|^2)^2} \dif z^{\mu} \wedge \dif \bar{z}^{\nu}.}
Here and hereafter, the Einstein sum convention is assumed.
Also we defined $|z|^2 := z^{\mu} \bar{z}^{\mu}$.
We can see that $\omega$ is a non-degenerate form.
Thus, $\omega$ is a symplectic structure on $CP^n$ and the local function $K=\log(1+|z|^2)$ satisfying (\ref{K on CP^n}) is called the K\"{a}hler potential.
We now define a standard almost complex structure $J$ by $J(\partial_{\mu}) = \im \partial_{\mu}, \ J(\partial_{\bar \mu}) = -\im \partial_{\bar \mu}$, where $\partial_{\mu} = \partial/ \partial z^{\mu}$ and $\partial_{\bar{\mu}} = \partial/ \partial \bar{z}^{\mu}$.
Then, the compatible metric $g(\cdot ,\cdot) := \omega(\cdot,J\cdot)$ is of the form
\eq{\label{metric} g = g_{\mu\bar{\nu}} \dif z^{\mu}\otimes \dif \bar{z}^{\nu} + g_{\bar {\nu} \mu} \dif \bar{z}^{\nu} \otimes \dif z^{\mu}.}
The components of the metric are given by
\eq{\label{metric component} g_{\mu\bar{\nu}} = g_{\bar {\nu} \mu} = \frac{(1+ |z|^2) \delta_{\mu\nu} - \bar{z}^{\mu} z^{\nu}}{(1+ |z|^2)^2} .}
This metric is called the Fubini-Study metric.
The volume element is given by
\eq{\sqrt{\det g} = (1+ |z|^2)^{-n-1}}
and the inverse metric is given by
\eq{\label{metric inverse} g^{\mu\bar{\nu}} = g^{\bar {\nu} \mu} = (1+ |z|^2) (\delta_{\mu\nu} + z^{\mu} \bar{z}^{\nu}). }
The triple $(\omega,g,J)$ gives the K\"ahler structure of $CP^n$.

Let us discuss the isometric embedding of $CP^n$ into $\mathbb{R}^{n^2 + 2n}$.
Let us consider a particular representative of homogeneous coordinate $\zeta = (\zeta^1, \zeta^2, \cdots ,\zeta^{n+1})$ such that $|\zeta|^2 = 1$.
On the patch $U_{n+1}$, for example, it is related to the inhomogeneous coordinate $z$ by
\eq{
\label{zeta}
\zeta = \frac{(z^1, z^2, \cdots, z^n , 1)^{\mathrm{T}}}{\sqrt{1+|z|^2}} \in \mathbb{C}^{n+1},
}
where we fix the phase of $\zeta$ so that $\zeta^{n+1}$ is a positive real number.
The rank 1 hermitian projection $P_{\zeta} = \zeta \zeta^{\dagger}$ can be expanded as
\eq{ \label{zeta and generators}
P_{\zeta} = \frac{1}{n+1} I_{n+1} - \sqrt{2} X^A T_A.
}
Here, $\{T_A\}_{I=1}^{n^2 + 2n}$ are Hermitian generators of $\mathrm{SU}(n+1)$ in fundamental representation satisfying
\eq{
\label{T_A algebra}
T_A T_B = \frac{1}{2(n+1)} \delta_{AB} I_{n+1} + \frac{1}{2}(d_{ABC} + \im f_{ABC})T_C.
}
$d_{ABC}$ and $f_{ABC}$ are the completely symmetric and anti-symmetric structure constants, respectively.
From the fact that $P_{\zeta}$ is a projector, the real coefficients $\{X^A\}_{A =1}^{n^2 +2n}$ satisfy
\eq{X^A X^A = \frac{n}{n+1}, \quad d_{ABC} X^A X^B + \sqrt{2}\left( \frac{n-1}{n+1} \right)X^C =0.}
A straightforward calculation shows that the Fubini-Study metric (\ref{metric}) can be written as
\eq{\dif s^2 = \tr( \dif P_{\zeta} \dif P_{\zeta}) = 2\tr(\dif X^A T_A \dif X^B T_B) = \dif X^A \dif X^A .}
Therefore, $\{X^A\}_{A =1}^{n^2 +2n}$ are isometric embedding functions.
$X^A$ can be also written as
\eq{\label{X^A}
X^A = - \sqrt{2}\zeta^{\dagger} T_A \zeta.}

Let us consider an action 
\eq{\label{SU(N) action} \zeta \mapsto U \zeta,}
for $U \in \mathrm{SU}(n+1)$.
This transformation leaves the metric invariant and hence is an isometry of $CP^n$.
Since $T_A$ is an invariant tensor of $\mathrm{SU}(n+1)$, the embedding functions $X^A$ transforms as the adjoint representation of $\mathrm{SU}(n+1)$.

Finally, let us consider the prequantum line bundle over $CP^n$.
One can construct $L$ as a dual bundle of the tautological line bundle over $CP^n$.
The curvature of $L$ is (\ref{k}) with $k=1$.
One can check that the integral of $\im R^L/2\pi$ over any 2-cycle is equal to $1$ as follows.
Since the rank of the second homology group of $CP^n$ is $1$, there is only one independent two-cycle.
Let us take a particular two-cycle $CP^1 = \{[Z^1,Z^2,0,\cdots,0]\} \subset CP^n$.
The symplectic form in this two-cycle is $\omega = \im \frac{\dif z \wedge \dif \bar{z}}{1+|z|^2}$, where $z=Z^1/Z^2$.
Then, it is easy to show that
\eq{ \frac{\im}{2\pi}\int_{CP^1} R^L = \frac{1}{2\pi}\int_{CP^1} \omega = 1.}

\subsection{Zero modes of the Dirac operator on $CP^n$}
\label{3.2 CP^n zeromode}

In this subsection, we construct a complete orthonormal basis of the Dirac zero modes on $CP^n$.

Let $D^{(p)}$ be a twisted Dirac operator on $\Gamma(S_c \otimes L^{\otimes p})$.
We take a specific representation of the gamma matrices given in (\ref{recursion}).
As shown in Appendix \ref{Vanishing}, the Dirac operator zero mode $\psi^{(p)} \in \Gamma(S_c \otimes L^{\otimes p})$ has only one spinor component $\psi^{(p)} = f^{(p)} \ket{+}^{\otimes n}$.
Here, $f^{(p)} \in \Gamma(L^{\otimes p})$ and $\ket{+}$ is two-dimensional spinor $(1,0)^{\mathrm T}$.
As shown in Appendix \ref{simplification}, the zero mode equation $D^{(p)}\psi^{(p)}=0$ is simplified to
\eq{ \left( \partial_{\bar{\mu}} +  p A^L_{\bar{\mu}} \right) f^{(p)} = 0.}
Plugging $K=\log (1+|z|^2)$ and $k=1$ into (\ref{A^L from K}), one finds
\eq{ p A^L_{\bar{\mu}} = \frac{p z^{\mu}}{2(1+|z|^2)}.}
Thus, the zero mode equation becomes
\eq{ \left(\partial_{\bar{\mu}} +\frac{pz^{\mu}}{2(1+|z|^2)} \right) f^{(p)} = 0,}
and the general solution to this equation is
\eq{ f^{(p)} = (1+|z|^2)^{-p/2}\phi(z),}
where $\phi(z)$ is an arbitrary holomorphic function.

Now, let us consider the norm of the zero modes.
Since any holomorphic function can be expanded in Taylor series around $z=0$, let us consider a function $\phi_{s}(z) := (z^1)^{s_1} (z^2)^{s_2} \cdots (z^n)^{s_n}$, where $s = (s_1,s_2,\cdots,s_n) \in (\mathbb{Z}_{\ge0})^n$ and check whether the zero mode $\psi^{(p)}_{s} = (1+|z|^2)^{-p/2}\phi_{s}\ket{+}^{\otimes n}$ is normalizable or not.
In Appendix \ref{integral formula}, we show that the norm
\eq{
\label{cp^n normalization}
|\psi^{(p)}_{s}|^2 = \int_{CP^n} \mu \, \frac{ |z^1|^{2s_1} |z^2|^{2s_2} \cdots |z^n|^{2s_n}}{(1+|z|^2)^{p}}}
is convergent if and only if $\sum_{i=1}^n s_i < p +1$ is satisfied.
It is shown in Appendix \ref{integral formula} that a complete orthonormal basis of $\mathrm{Ker} \, D^{(p)}$ can be chosen as
\als{
\label{zero modes in z}
&\psi^{(p)}_{s} = (I_{s,p})^{-1/2}(1+|z|^2)^{-p/2} (z^1)^{s_1} (z^2)^{s_2} \cdots (z^n)^{s_n}\ket{+}^{\otimes n},\\
& {}^{\forall} i \in \{1,2,\cdots,n\}: \ s_i \in \mathbb{Z}_{\ge 0} \quad \text{s.t.} \quad  \sum_{i=1}^n s_i \le p,
}
where $I_{s,p}$ is given in (\ref{I_s,p}).

There is another expression of (\ref{zero modes in z}) in terms of the normalized inhomogeneous coordinate $\zeta$ given in (\ref{zeta}).
The orthonormal basis (\ref{zero modes in z}) can be written as
\eq{
\psi^{(p)}_{\bm{\alpha}_p} = c^{(p)}_{\bm{\alpha}_p} f^{(p)}_{\bm{\alpha}_p} \ket{+}^{\otimes n},
}
where
\eq{
f^{(p)}_{\bm{\alpha}_p} = \zeta^{\alpha_1} \zeta^{\alpha_2} \cdots \zeta^{\alpha_p},
}
and the collective index $\bm{\alpha}_p = (\alpha_1,\alpha_2,\cdots,\alpha_p)$ is an element of 
\eq{ \Sigma_p = \{1,2,\cdots,n+1\}^p/\text{permutation}.}
The normalization factor $c^{(p)}_{\bm{\alpha}_p}$ is given by
\eq{\label{normalization constant}
c^{(p)}_{\bm{\alpha}_p} = \sqrt{\frac{(p+n)!}{(2\pi)^n \prod_{i=1}^{n+1} n_i(\bm{\alpha}_p)!} },}
where $n_{i}(\bm{\alpha}_p)$ is the number of components of $\bm{\alpha}_p$ equal to $i$.

The dimension of the $\mathrm{Ker} \, D^{(p)}$ is
\eq{ 
\label{dim ker D}
\dim \mathrm{Ker} \, D^{(p)} = \frac{(n+p)!}{n! p!},
}
which is the  number of independent symmetric polynomials of degree $p$ with $n$ variables.
(\ref{dim ker D}) can also be understood from representation theory of $\mathfrak{su}(n+1)$.
Let $V_{(d_1,d_2,\cdots,d_n)}$ be an irreducible representation of $\mathfrak{su}(n+1)$ with Dynkin index $(d_1,d_2,\cdots,d_n)$.
From (\ref{SU(N) action}), one can see that $\zeta$ is in the representation space $V_{(1,0,\cdots,0)}$, which implies that the set of all symmetric polynomials of $\zeta^i$ of degree $p$ is isomorphic to the representation space $V_{(p,0,\cdots,0)}$.
Thus, we have
\eq{ \mathrm{Ker} \, D^{(p)} = V_{(p,0,\cdots,0)}.}
According to the hook length formula, the dimension of $V_{(p,0,\cdots,0)}$ is indeed equal to $\frac{(n+p)!}{n! p!}$.
This viewpoint in terms of representation theory will also play a very important role in the following discussions.

As calculated in \cite{Dolan:2002ck}, one can also obtain $ \dim \mathrm{Ker} \, D^{(p)}$ from the index theorem.
Since the vanishing theorem holds, we have $\dim \mathrm{Ker} \, D^{(p)} = \mathrm{Ind} D^{(p)}$.
Then, from the index theorem, we obtain
\eq{
\label{CP^n index theorem}
\dim \mathrm{Ker} \, D^{(p)} = \int_{CP^n} \mathrm{Td}(T^{(1,0)}CP^n) \wedge \mathrm{ch}(L^{\otimes p}),}
where $\mathrm{Td}$ and $\mathrm{ch}$ stand for the Todd class and Chern character, respectively.
For $CP^n$, we have
\footnote{We sometimes write $\alpha^{n}$ for $\alpha^{\wedge n}$ for any differential form $\alpha$. The exponential of a differential form $\alpha$ is defined as $\e^{\alpha} = \sum_{k=0}^{\infty} \frac{\alpha^{\wedge k}}{k!}$.}
\eq{ \mathrm{Td}(T^{(1,0)}CP^n) = \left( \frac{\omega/2\pi}{1 - \e^{-\omega/2\pi}} \right)^{n+1} , \quad \mathrm{ch}(L^{\otimes p}) = \e^{p \omega/2\pi}.}
The coefficient of the term proportional to $(\omega/2\pi)^{\wedge n}$ in the integrand of (\ref{CP^n index theorem}) can be evaluated using the residue theorem:
\eq{C_{p,n} := \frac{1}{2\pi \im} \oint \frac{\dif z}{z^{n+1}} \left( \frac{z}{1 - \e^{-z}} \right)^{n+1} \e^{p z} = \frac{1}{2\pi \im} \oint \dif z \frac{\e^{p z}}{( 1 - \e^{-z})^{n+1}},}
where the integration contour is a counterclockwise loop enclosing the origin $z=0$.
By integrating by parts, one can verify
\eq{C_{p,n} = \frac{p+1}{n} C_{p+1,n-1} = \cdots = \frac{(n+p)!}{n!p!} C_{n+p,0} = \frac{(n+p)!}{n!p!}.}
To obtain the last equality, we use
\eq{C_{n+p,0} = \frac{1}{2\pi \im} \oint \dif z \frac{\e^{(n+p) z}}{1 - \e^{-z}} = \frac{1}{2\pi \im} \oint \dif z \frac{\e^{(n+p) z}}{z} 
\left( \sum_{l=0}^{\infty} \frac{(-z)^l}{(l+1)!} \right)^{-1} = 1.}
Using the result of Appendix \ref{integral formula}, we have $\int_{CP^n} \left( \frac{\omega}{2\pi}\right)^{\wedge n} = 1$ and we therefore obtain
\eq{ \dim \mathrm{Ker} \, D^{(p)} =  \int_{CP^n} C_{p,n} \left( \frac{\omega}{2\pi}\right)^{\wedge n} = C_{p,n} = \frac{(n+p)!}{n!p!}.}

\subsection{Matrix regularization of embedding functions}

We will show that the embedding functions $\{X^A\}_{A=1}^{n^2 +2n}$ defined in (\ref{X^A}) are mapped to
\eq{ \label{embedding regularization}
T_p(X^A) = \frac{\sqrt{2}}{p+n+1} L^{(p)}_A.}
Here, $\{L^{(p)}_A\}_{A=1}^{n^2+2n}$ are generators of $\mathrm{SU}(n+1)$ in the irreducible representation with Dynkin index $(p,0,\cdots,0)$ satisfying
\eq{ \label{properties of L^A}
[L^{(p)}_A,L^{(p)}_B]  = \im f_{ABC}L^{(p)}_C, \quad (L_A^{(p)})^2 = \frac{np(p+n+1)}{2(n+1)} I.}

Let $\bm{\alpha}_p, \bm{\beta}_p \in \Sigma_p$ be collective indices labelling the orthonormal basis of $\mathrm{Ker} \, D^{(p)}$.
From (\ref{X^A}), the Toeplitz operator $T_p (X^A)$ is given by
\als{ 
\label{T(X^A) cp^n 1}
T_p (X^A)_{\bm{\alpha}_p, \bm{\beta}_p} :=& \int_{CP^n} \mu \, (\psi^{(p)}_{\bm{\alpha}_p})^{\dagger} X^A \psi^{(p)}_{\bm{\beta}_p} \\
=& -\sqrt{2} \sum_{i,j=1}^{n+1}(T_A)_{ij} c^{(p)}_{\bm{\alpha}_p} c^{(p)}_{\bm{\beta}_p}  \int_{CP^n} \mu \, (f^{(p)}_{\bm{\alpha}_p})^* f^{(p)}_{\bm{\beta}_p} \zeta^{j} \bar{\zeta}^{i} \\
=& -\sqrt{2} \sum_{i,j=1}^{n+1}(T_A)_{ij} c^{(p)}_{\bm{\alpha}_p} c^{(p)}_{\bm{\beta}_p}  \int_{CP^n} \mu \, (f^{(p+1)}_{\bm{\alpha}_p \oplus i})^* f^{(p+1)}_{\bm{\beta}_p \oplus j}.}
Here, we introduced the notation $\bm{\alpha}_p \oplus \bm{\gamma}_l = (\alpha_1,\alpha_2,\cdots, \alpha_p ,\gamma_1, \gamma_2,\cdots,\gamma_l) \in \Sigma_{p+l}$ for $\bm{\alpha}_p = (\alpha_1,\alpha_2,\cdots, \alpha_p) \in \Sigma_p$ and $\bm{\gamma}_l = (\gamma_1, \gamma_2,\cdots,\gamma_l) \in \Sigma_l$.
Using the orthonormality condition, we have
\als{ 
\label{T(X^A) CP^n}
T_p (X^A)_{\bm{\alpha}_p, \bm{\beta}_p} &= -\sqrt{2} \sum_{i,j=1}^{n+1}(T_A)_{ij} \left(\frac{c^{(p)}_{\bm{\alpha}_p}}{c^{(p+1)}_{\bm{\alpha}_p \oplus i}} \right)^2 \frac{c^{(p)}_{\bm{\beta}_p}}{c^{(p)}_{\bm{\alpha}_p}} \delta_{\bm{\alpha}_p \oplus i,\bm{\beta}_p \oplus j}\\
&= - \frac{\sqrt{2}}{p+n+1} \frac{c^{(p)}_{\bm{\beta}_p}}{c^{(p)}_{\bm{\alpha}_p}} \sum_{i,j=1}^{n+1}(T_A)_{ij} (n_{i}(\bm{\alpha}_p) +1)  \delta_{\bm{\alpha}_p \oplus i,\bm{\beta}_p \oplus j}.}
The Kronecker delta $\delta_{\bm{\alpha}_p, \bm{\beta}_p}$ is defined by
\eq{\delta_{\bm{\alpha}_p, \bm{\beta}_p} = \begin{cases}
    1 & (\bm{\alpha}_p = \bm{\beta}_p)\\
    0 & (\bm{\alpha}_p \neq \bm{\beta}_p)
\end{cases},}
 and we used
\eq{ \label{formula 3}
\frac{c^{(p+1)}_{\bm{\alpha}_p \oplus i}}{c^{(p)}_{\bm{\alpha}_p}}  = \sqrt{\frac{p+n+1}{n_{i}(\bm{\alpha}_p) +1}},}
in the second equality.

Second, let us define
\eq{ L^{(p)}_A := \frac{p+n+1}{\sqrt{2}}T_p(X^A),\quad
\mathcal{L}_A^{(0)} X^B := -\frac{\im}{\sqrt{2}} \{X^A,X^B\},
}
where the Poisson bracket $\{X^A,X^B\}$ is given in (\ref{Poisson bracket CP^n}).
From (\ref{L_A correspondence}), they satisfy
\eq{
\label{correspondence 1}
[L^{(p)}_A, T_p (X^B)] = T_p (\mathcal{L}_A^{(0)} X^B).}
By using (\ref{correspondence 1}) and (\ref{Poisson bracket X^A}), we find
\eq{ [L^{(p)}_A, L^{(p)}_B] = \frac{p+n+1}{\sqrt{2}} T_p (\mathcal{L}_A^{(0)} X^B) = \im f_{ABC}\frac{p+n+1}{\sqrt{2}} T_p (X^C) = \im f_{ABC}L^{(p)}_C.}
This shows that $\{L^{(p)}_A\}_{A=1}^{n^2 +2n}$ is $\mathrm{SU}(n+1)$ generators in some representation.
To identify the representation, let us calculate the quadratic Casimir.
From (\ref{T(X^A) CP^n}), we obtain
\eq{ (L_A^{(p)})^2_{\bm{\alpha}_p, \bm{\beta}_p} = \frac{c^{(p)}_{\bm{\beta}_p}}{c^{(p)}_{\bm{\alpha}_p}} \sum_{i,j,i',j'=1}^{n+1} (n_{i}(\bm{\alpha}_p) +1) (n_{j'}(\bm{\alpha}_p)+1+\delta_{i,j'} - \delta_{j,j'} ) (T_A)_{ij}(T_A)_{j'i'} \delta_{\bm{\alpha}_p \oplus i \oplus j',\bm{\beta}_p \oplus i' \oplus j}.
}
Using the Fierz identity
\eq{ 
\label{Fierz}
(T_A)_{ij}(T_A)_{j'i'} = \frac{1}{2} \left( \delta_{i,i'} \delta_{j,j'} - \frac{1}{n+1} \delta_{ij} \delta_{i'j'}\right),}
we obtain
\eq{
\label{casimir}
(L_A^{(p)})^2_{\bm{\alpha}_p, \bm{\beta}_p} = \frac{np(p+n+1)}{2(n+1)} \delta_{\bm{\alpha}_p, \bm{\beta}_p}.
}
This is exactly the quadratic Casimir eigenvalue of representation $(p,0,\cdots,0)$ and therefore $\{L^{(p)}_A\}_{A=1}^{n^2 +2n}$ is in the irreducible representation $(p,0,\cdots,0)$.

\subsection{Laplace operator on $\Gamma(L^{\otimes q})$}

Consider a Laplace operator on $\Gamma(L^{\otimes q})$
\eq{ \Delta^{(q)} = - g^{\mu \nu} \nabla_{\mu} \nabla_{\nu} = - \{X^A, \{X^{A},f^{(q)}\}\},}
for $f^{(q)} \in \Gamma(L^{\otimes q})$.
Here, $\{ \cdot , \cdot\}$ is a generalized Poisson bracket defined in (\ref{generalized Poisson bracket}) and $\{X^A\}_{A=1}^{n^2 +2n}$ is isometric embedding functions.
Let us also define differential operators $\{ \mathcal{L}^{(q)}_A \}_{A=1}^{n^2 +2n}$ on $\Gamma(L^{\otimes q})$ by
\eq{
\label{generators on sections}
\mathcal{L}^{(q)}_A f^{(q)} := \frac{1}{\sqrt{2}}\left(- \im \{X^{A},f^{(q)}\} +q X^A f^{(q)} \right).
}
As shown in Appendix \ref{differential operator commutation relation}, they satisfy commutation relations of generator of $\mathrm{SU}(n+1)$:
\eq{
\label{algebra of generators on sections}
[\mathcal{L}^{(q)}_A,\mathcal{L}^{(q)}_B] = \im f_{ABC}\mathcal{L}^{(q)}_C.
}
By a straightforward calculation, we can derive
\eq{
\label{Laplacian and Casimir}
\Delta^{(q)}= 2(\mathcal{L}_A^{(q)})^2 - \frac{q^2 n}{n+1}.
}
Thus, the eigenvalue of $\Delta^{(q)}$ is given by $2E - \frac{q^2 n}{n+1}$, where $E$ is an eigenvalue of $(\mathcal{L}_A^{(q)})^2$.

Let us evaluate the eigenvalues of $(\mathcal{L}_A^{(q)})^2$.
To do this, let us consider how one can write elements of $\Gamma(L^{\otimes q})$ in term of local coordinates.
Remind that in the overlapping patch $U_{\alpha} \cap U_{\beta}$, $A^L$ transforms as
\eq{ A^L (z_{(\alpha)}) =  A^L (z_{(\beta)}) - \dif \lambda(z_{(\beta)}),}
where
\eq{\lambda(z_{(\beta)}) = -\frac{1}{2} \left[\log(\frac{Z^{\alpha}}{Z^{\beta}})- \log(\frac{\bar{Z}^{\alpha}}{\bar{Z}^{\beta}})\right].}
Here, $Z=(Z^1,Z^2,\cdots,Z^{n+1})$ is the homogeneous coordinates of $CP^n$.
Hence, any element $f^{(q)} \in \Gamma(L^{\otimes q})$ should transforms as
\eq{ f^{(q)}(z_{(\alpha)}) = \e^{q \lambda( z_{(\beta)})} f^{(q)}(z_{(\beta)}) = \left(\frac{Z^{\alpha}}{Z^{\beta}}\right)^{-\frac{q}{2}} \left(\frac{\bar{Z}^{\alpha}}{\bar{Z}^{\beta}}\right)^{\frac{q}{2}}f^{(q)}(z_{(\beta)}).}
Thus, we can choose a basis of $\Gamma(L^{\otimes q})$ as
\eq{
(Z^{\mu} \bar{Z}^{\mu})^{-k - \frac{q}{2}} Z^{\sigma_1} Z^{\sigma_2} \cdots Z^{\sigma_{k+q}} \bar{Z}^{\tau_1}
\bar{Z}^{\tau_2} \cdots \bar{Z}^{\tau_{k}},}
where $k \in \mathbb{Z}_{\ge0}$.
With the normalized homogeneous coordinates $\zeta$ given in (\ref{zeta}), we define a basis of $\Gamma(L^{\otimes q})$ as
\eq{
\label{basis of monopole section}
f_{\bm{\sigma}_{k+q}, \bm{\tau}_{k}}^{(q)}(z) := \zeta^{\sigma_1} \zeta^{\sigma_2} \cdots \zeta^{\sigma_{k+q}} \bar{\zeta}^{\tau_1}
\bar{\zeta}^{\tau_2} \cdots \bar{\zeta}^{\tau_{k}},}
From (\ref{zeta and generators}), one can see that $\zeta$ and $\bar{\zeta}$ are in the representation spaces $V_{(1,0,\cdots,0)}$ and $V_{(1,0,\cdots,0)}^*$, respectively, where $V_{(1,0,\cdots,0)}^*$ is the complex conjugate representation space of $V_{(1,0,\cdots,0)}$.
This implies that the set of all polynomials of $\zeta^i, \bar{\zeta}^j$ of degree $(k+q,k)$ denoted by $\mathrm{Pol}_{k+q,k}(\zeta,\bar{\zeta})$ is isomorphic to $V_{(k+q,0,\cdots,0)} \otimes V_{(k,0,\cdots,0)}^*$, because of the symmetric index structure of the polynomials.
Using the irreducible decomposition $V_{(k+q,0,\cdots,0)} \otimes V^*_{(k,0,\cdots,0)} = \bigoplus_{i=0}^{k} V_{(i+q,0,\cdots,0,i)}$, we have
\eq{ \label{charged field decomposition}
\Gamma(L^{\otimes q}) = \bigoplus_{k=0}^{\infty}\mathrm{Pol}_{k+q,k}(\zeta,\bar{\zeta}) = \bigoplus_{k=0}^{\infty} \left( V_{(k+q,0,\cdots,0)} \otimes V_{(k,0,\cdots,0)}^* \right) = \bigoplus_{k=0}^{\infty} V_{(k+q,0,\cdots,0,k)}. }
The eigenvalues of $(\mathcal{L}_A^{(q)})^2$ are those of the quadratic Casimir for the representations $(k+q,0,\cdots,0,k)$, which are given by
\eq{
\label{E_k CP^n}
E_k = \frac{1}{2}\left( (k+q)(k+n) + k (k+q+n) + \frac{nq^2}{n+1}\right).}

We can find eigenvectors of $(\mathcal{L}_A^{(q)})^2$ from the similar group theoretic correspondence.
The eigenvectors corresponding to $V_{(k+q,0,\cdots,0,k)}$ are
\eq{
\label{f_k,I}
f^{(q)}_{k,I} (z) = \sum_{\bm{\sigma}_{k+q}, \bm{\tau}_k} c^{(q)}_{I, \bm{\sigma}_{k+q}, \bm{\tau}_k} f_{\bm{\sigma}_{k+q}, \bm{\tau}_{k}}^{(q)}(z) = \sum_{\bm{\sigma}_{k+q}, \bm{\tau}_k} c^{(q)}_{I, \bm{\sigma}_{k+q}, \bm{\tau}_k} \zeta^{\sigma_1}
\zeta^{\sigma_2} \cdots \zeta^{\sigma_{k+q}} \bar{\zeta}^{\tau_1} \bar{\zeta}^{\tau_2} \cdots \bar{\zeta}^{\tau_k},}
where $c^{(q)}_{I, \bm{\sigma}_{k+q}, \bm{\tau}_k} := (c^{(q)}_I)_{\sigma_1 \cdots\sigma_{k+q},\tau_1 \cdots \tau_k}$ is a coefficient tensor which is completely symmetric in $\sigma$ and $\tau$, respectively, and traceless under any contraction between $\sigma_a$ and $\tau_b$.
The index $I$ labels different weights of $V_{(k+q,0,\cdots,0,k)}$
\footnote{For example, for $n=1$, we can take eigenvalues of $\mathcal{L}^{(q)}_3$ as the index $I$.}.
We also choose $c^{(q)}_{I, \bm{\sigma}_{k+q}, \bm{\tau}_k}$ such that
\eq{
\label{CP^n monopole bundle normalization}
(f^{(q)}_{k,I},f^{(q)}_{k',I'}) := \int_{CP^n} \mu \,(f^{(q)}_{k,I})^{*} f^{(q)}_{k',I'} = \delta_{k,k'} \delta_{I,I'}. 
}

In Appendix \ref{cp^n laplacian eigenvalues}, we show a direct computation of (\ref{E_k CP^n}).

\subsection{Matrix regularization of $\Gamma(L^{\otimes q})$ and the Laplace operator}

In this subsection, we explicitly evaluate the Toeplitz operator for a complete basis of $\Gamma(L^{\otimes q})$ given by the eigenfunctions of $\Delta^{(q)}$ and discuss the matrix Laplacian.

Let us consider a matrix regularization of $\Gamma(L^{\otimes q})$ by
\eq{ T_p (f^{(q)}) = \Pi^{(p+q)} f^{(q)} \Pi^{(p)}, \quad (f^{(q)} \in \Gamma(L^{\otimes q})) }
where $\Pi^{(p)}: \Gamma(S_c \otimes L^{\otimes p}) \to \mathrm{Ker} \, D^{(p)}$ is the projection.
As discussed in the previous subsection, we can choose a complete basis of $\Gamma(L^{\otimes q})$ by
\eq{ f_{\bm{\sigma}_{k+q}, \bm{\tau}_{k}}^{(q)}(z) = \zeta^{\sigma_1}
\zeta^{\sigma_2} \cdots \zeta^{\sigma_{k+q}} \bar{\zeta}^{\tau_1} \bar{\zeta}^{\tau_2} \cdots \bar{\zeta}^{\tau_k}.}
Then, the matrix regularization of $f_{\bm{\sigma}_{k+q}, \bm{\tau}_{k}}^{(q)}$ is given by
\eq{ 
\label{T(f)}
T_p (f_{\bm{\sigma}_{k+q}, \bm{\tau}_{k}}^{(q)})_{\bm{\alpha}_{p+q}, \bm{\beta}_p} := \int_{CP^n} \mu \, (\psi^{(p+q)}_{\bm{\alpha}_{p+q}})^{\dagger} f^{(q)}_{\bm{\sigma}_{k+q}, \bm{\tau}_{k}} \psi^{(p)}_{\bm{\beta}_p} =  \frac{c^{(p+q)}_{\bm{\alpha}_{p+q}} c^{(p)}_{\bm{\beta}_p}}{\left(c^{(p+q+k)}_{\bm{\alpha}_{p+q} \oplus \bm{\tau}_k} \right)^2} \delta_{\bm{\alpha}_{p+q} \oplus \bm{\tau}_k, \bm{\beta}_{p} \oplus \bm{\sigma}_{k+q}}.
}

From (\ref{fuzzy matrix laplacian}) and $\hbar_{p} = p^{-1}$, we define a Laplace operator acting on $T_p (f^{(q)})$ by
\eq{\hat{\Delta}(T_p (f^{(q)})) = p^2 [T(X^A),[T(X^A),T_p (f^{(q)})]].}
Using (\ref{embedding regularization}) and (\ref{properties of L^A}), we have
\eq{\hat{\Delta}(T_p (f^{(q)})) = \frac{2p^2}{(p+q+n+1)(p+n+1)} \left((L_A \circ)^2 - \frac{ q^2 n}{2(n+1)}\right)T_p (f^{(q)}),
}
where we defined $L_A \circ T_p (f^{(q)}) := L^{(p+q)}_A T_p (f^{(q)}) - T_p (f^{(q)}) L^{(p)}_A$.
The operation $L_A \circ$ satisfies
\eq{ [L_A \circ,L_B \circ] = \im f_{ABC} L_C \circ,}
and hence they are representations of the generators of $\mathrm{SU}(n+1)$.
Their representation space is
\eq{V_{(p+q,0,\cdots,0)} \otimes V^*_{(p,0,\cdots,0)} = \bigoplus_{k=0}^{p} V_{(k+q,0,\cdots,0,k)}.}
This is a similar decomposition to (\ref{charged field decomposition}) except for the cut off $p$.
From this, we see that the eigenmatrices of $\hat{\Delta}$ are in the irreducible representation $V_{(k+q,0,\cdots,0,k)}$ and the eigenvalue of $\hat{\Delta}$ is given by
\eq{
\frac{2p^2}{(p+q+n+1)(p+n+1)}\left(E_k - \frac{ q^2 n}{2(n+1)}\right) = 2E_k - \frac{q^2 n}{n+1} + O(p^{-1})
}
for $k=1,2,\cdots,p$, where $E_k$ is given by (\ref{E_k CP^n}). 
This shows that the spectrum of the matrix Laplacian $\hat{\Delta}$ is the truncated version of the spectrum of the Bochner Laplacian $\Delta$ up to a correction of order $O(1/p)$.

More explicitly, we can show the stronger identity
\eq{
\label{L_A correspondence}
T_p (\mathcal{L}_A^{(q)} f^{(q)}) = L_A \circ T_p (f^{(q)})}
for any $f^{(q)} \in \Gamma(L^{\otimes q})$.
This is shown in Appendix \ref{su(n) generator correspondence}.
From this identity, we can easily derive the correspondence of eigenvalues or eigenvectors we discussed above.
Note that $T_p (f^{(q)}_{k,I})$ can be written as
\als{ T_p (f^{(q)}_{k,I})_{\bm{\alpha}_{p+q}, \bm{\beta}_p} =  \int_{CP^n} \mu \, (\psi^{(p+q)}_{\bm{\alpha}_{p+q}})^{\dagger} f^{(q)}_{k,I} \psi^{(p)}_{\bm{\beta}_{p}} = c^{(p+q)}_{\bm{\alpha}_{p+q}} c^{(p)}_{\bm{\beta}_{p}} (f^{(q)}_{\bm{\alpha}_{p+q}, \bm{\beta}_p}, f^{(q)}_{k,I}),
}
where $(\cdot,\cdot)$ is the inner product defined in (\ref{CP^n monopole bundle normalization}).
Since $f^{(q)}_{\bm{\alpha}_{p+q}, \bm{\beta}_p}$ can be expanded by the orthonormal basis $f^{(q)}_{k',I'}$ for $k'\le p$, we find $T_p (f^{(q)}_{k,I}) =0$ for $k>p$.
For $k \le p$, (\ref{L_A correspondence}) implies that $f^{(q)}_{k,I}$ and $T_p (f^{(q)}_{k,I})$ both have exactly the same Casimir eigenvalues and weights.
For the quadratic Casimir, we have
\eq{(L_A \circ)^2 T_p (f^{(q)}_{k,I}) = T_p ((\mathcal{L}_A^{(q)})^2 f^{(q)}_{k,I}) = E_k T_p (f^{(q)}_{k,I}),}
and we can see that the eigenvalues of $(L_A \circ)^2$ are $\{E_k\}_{k=0}^{p}$ as expected.

To see the correspondence of the trace and the integral (\ref{trace property}), let us calculate the Frobenius inner product,
\eq{
\label{Frobenius CP^n}
(T_p(f^{(q)}_{k,I}),T_p(f^{(q)}_{k',I'})):= \Tr [T_p(f^{(q)}_{k,I})^{\dagger} T_p(f^{(q)}_{k',I'})].
}
For $k,k' \le p$, (\ref{Frobenius CP^n}) is nonvanishing only if $T_p(f^{(q)}_{k,I})$ and $T_p(f^{(q)}_{k',I'})$ belong to the same representation having the same weights.
Thus, we have
\eq{
\label{orthogonality of matrix}
(T_p(f^{(q)}_{k,I}),T_p(f^{(q)}_{k',I'})) \propto \delta_{k,k'} \delta_{I,I'}.
}
More explicitly, we can show
\eq{ \label{cp^n inner product correspondence}
(T_p(f^{(q)}_{k,I}),T_p(f^{(q)}_{k',I'})) = \frac{(p+q+n)!(p+n)!}{(2\pi)^n (p-k)!(p+q+k+n)!} \delta_{k,k'} \delta_{I,I'} .
}
See Appendix \ref{cp^n inner product eigenfunctions} for the proof.
For finite $k$ and $k'$, we have the large $p$ expansion,
\eq{ (T_p(f^{(q)}_{k,I}),T_p(f^{(q)}_{k',I'})) 
= \frac{p^n}{(2\pi)^n} \delta_{k,k'} \delta_{I,I'} + O(p^{n-1}),
}
which is consistent with (\ref{CP^n monopole bundle normalization}) through the correspondence for the trace and integral (\ref{trace property}).

\section{Fuzzy $T^{2n}$}
In this section, we consider a Berezin-Toeplitz quantization of monopole bundle over a torus $T^{2n} \simeq (S^1)^{2n}$ \cite{Adachi:2020asg}.
In subsection 4.1, we define a torus $T^{2n}$ and describe basic properties. 
In subsection 4.2, we explicitly construct a complete orthonormal basis of the kernel of the Dirac operator. 
In subsection 4.3, we calculate Toeplitz operators of embedding functions. 
In subsection 4.4 and 4.5, we discuss the continuum Laplacian and the matrix Laplacian, respectively, for a monopole bundle
\footnote{In \cite{Adachi:2020asg}, the two-dimensional case is studied.
In this paper, we study its higher dimensional extension.}.

\subsection{Geometry of $T^{2n}$}

Let us consider the Euclidean space $\mathbb{R}^{2n}$ equipped with a flat metric.
We introduce an equivalent relation
\eq{ {}^{\forall} x=(x^1,x^2,\cdots,x^{2n}) \in \mathbb{R}^{2n} : \quad x^a \sim x^a + 2\pi l_a \quad (a=1,2,\cdots,2n),}
where $l_a$ are some positive constants. 
Under this identification, we define $2n$-dimensional torus $T^{2n}$ as a quotient space
\eq{T^{2n} = \mathbb{R}^{2n}/\sim.}
The flat metric and its associated K\"{a}hler form on $T^{2n}$ are given by
\eq{
g = \sum_{a=1}^{2n} \dif x^a \otimes \dif x^a,\quad
\omega = \sum_{m=1}^n \dif x^{2m-1} \wedge \dif x^{2m} = \im \, \dif z^{\mu} \wedge \dif \bar{z}^{\mu}.}
Here, the real and complex coordinates are related by $z^{\mu} = (x^{2\mu -1} + \im x^{2\mu })/\sqrt{2}$ for $\mu=1,2,\cdots,n$.
$T^{2n}$ is isometrically embedded in $\mathbb{R}^{4n}$ such that
\eq{ X^{2a-1} = l_a \cos (x^a/l_a), \quad X^{2a} = l_a\sin (x^a/l_a). \quad (a=1,2,\cdots,2n)}

Now, let us consider the bundle structures on $T^{2n}$.
Since $T^{2n}$ is a spin manifold, we can simply use the spin bundle $S$.
Since $T^{2n}$ is flat, the spin connection of $S$ is flat as well.
We also introduce the prequantum line bundle $L$.
The 2-cycles of $T^{2n}$ are simply $T^{2}$ and the curvature $R^L= - \im k \omega$ is non-vanishing on $T^2$ spanned by $(x^{2m-1},x^{2m})$ for $m=1,2,\cdots,n$.
Hence, the prequantization condition for $T^{2n}$ is satisfied for $k$ and $l_a$ such that
\eq{ \label{prequantization condition T^2n}
{}^{\forall} m \in \{1,2,\cdots,n\}: \quad q_m:=\frac{\im}{2\pi} \int_{T^2} R^L = 2\pi k l_{2m-1} l_{2m} \in \mathbb{N}.}
The condition is satisfied if and only if the ratio of areas $\frac{l_{2m-1} l_{2m}}{l_{2m'-1} l_{2m'}}$ is rational for any $m,m'$.

\subsection{Zero modes of the Dirac operator on $T^{2n}$}

In this subsection, we construct a complete orthonormal basis of the Dirac zero modes on $T^{2n}$ \cite{Tenjinbayashi:2005sy}.

Let $D^{(p)}$ be a twisted Dirac operator on $\Gamma(S \otimes L^{\otimes p})$.
By the same argument as in section \ref{3.2 CP^n zeromode}, the zero mode equation $D^{(p)}\psi^{(p)}=0$ for $\psi^{(p)} = f^{(p)} \ket{+}^{\otimes n}$ is simplified to
\eq{ \left( \partial_{\bar{\mu}} + p A^L_{\bar{\mu}} \right) f^{(p)} = 0.}
Here, $A^L$ can be chosen as
\eq{A^L = - \im k \sum_{m=1}^n x^{2m-1} \dif x^{2m} = - \, \frac{k}{2} ( z^{\mu} + \bar{z}^{\mu} )(\dif z^{\mu} - \dif \bar{z}^{\mu}).}
Thus, the zero-mode equation is
\eq{
\label{Df=0 torus}
\left( \partial_{\bar{\mu}} + \frac{kp}{2} (z^{\mu}+\bar{z}^{\mu}) \right) f^{(p)} = 0.}
We also have to pay attention to the boundary conditions.
Since $f^{(p)}(x)$ is a section of the nontrivial bundle $L^{\otimes p}$, $f^{(p)}(x)$ transforms under a coordinate change.
For $T^{2n}$, this property is described in terms of the boundary conditions as follows.
Consider the coordinate change $x^{2m} \mapsto x^{2m} + 2\pi l_{2m}$.
Under this change, the connection one-form $A^L(x)$ does not change and correspondingly the element of $\Gamma(L^{\otimes p})$ should be periodic under this coordinate shift for each $m$.
Similarly, under the coordinate change $x^{2m-1} \mapsto x^{2m-1} + 2\pi l_{2m-1}$, $A^L(x)$ transforms as $A^L(x) \mapsto A^L(x) - \dif \lambda(x)$ where $\lambda(x) =  \im 2\pi k l_{2m-1} x^{2m}$.
Correspondingly, $f^{(p)}$ should transforms as $f(x) \mapsto \e^{p\lambda(x)} f^{(p)}(x) = \e^{\im 2\pi kp l_{2m-1} x^{2m}} f^{(p)}(x)$ for each $m$.
These boundary conditions and the differential equation (\ref{Df=0 torus}) are closed on each $T^2$ with the coordinates $(x^{2m-1},x^{2m})$.
Hence, we can separate the variables and the general solution is
\eq{ 
\label{torus zero mode 1}
f^{(p)}(x) = \prod_{m=1}^n \left( \e^{-\frac{kp}{2} (x^{2m-1})^2 } \phi_{m}(x^{2m-1}+ \im x^{2m}) \right).}
The boundary conditions are now given by
\als{ 
\label{Torus BC}
&\phi_{m}(x^{2m-1}+ \im x^{2m} + \im 2\pi l_{2m}) = \phi_{m}(x^{2m-1}+ \im x^{2m}),\\
& \phi_{m}(x^{2m-1}+ \im x^{2m} + 2\pi l_{2m-1}) = \e^{- \im pq_m \tau_m} \e^{p q_m (x^{2m-1} + \im x^{2m})/l_{2m}} \phi_{m}(x^{2m-1}+ \im x^{2m}). }
Here, $\tau_m := \im l_{2m-1}/l_{2m}$ is the moduli parameter of the $m$-th $T^2$.
From the first condition, one can write
\eq{ \phi_{m} (x^{2m-1}+ \im x^{2m}) =  \sum_{s \in \mathbb{Z}} d_{s} \e^{s (x^{2m-1} + \im x^{2m})/l_{2m}},
}
for some complex constants $d_{s}$.
The second condition gives
\eq{ 
d_{s} = \e^{\im \pi(2s -pq_m)\tau_m} d_{s -pq_m}.}
To solve this recursion equation, let us write $s = pq_m l + i_m$ for $l \in \mathbb{Z}$ and $i_m \in \{0,1,\cdots,pq_m-1\}$.
Then, the solution is
\eq{d_{pq_m l + i_m} = c^{(p)}_{i_m} \e^{\im\pi \left(l + \frac{i_m}{pq_m} \right)^2 pq_m\tau_m},}
for some complex constants $c^{(p)}_{i_m}$.
Hence, there are $pq_m$ linearly independent solution to (\ref{Torus BC})
\eq{ \phi_{m} (x^{2m-1}+ \im x^{2m}) = \sum_{i_m = 0}^{pq_m -1} c^{(p)}_{i_m} \sum_{l \in \mathbb{Z}} \e^{\im\pi \left(l + \frac{i_m}{pq_m} \right)^2 pq_m\tau_m} \e^{\left(l + \frac{i_m}{pq_m} \right) \frac{pq_m}{l_{2m}}(x^{2m-1} + \im x^{2m})}.
}
Therefore, from (\ref{torus zero mode 1}), one can take a complete basis of the zero mode solutions as
\eq{ 
\label{torus zero mode 2}
f^{(p)}_{i}(x) = \prod_{m=1}^n f^{(p)}_{i_m} (x^{2m-1}, x^{2m}),}
where $i = (i_1,i_2,\cdots,i_n)$ and
\eq{ 
\label{torus zero mode 3}
f^{(p)}_{i_m} (x^{2m-1}, x^{2m}) := \left( \frac{kp}{4 \pi^3 l_{2m}^2}\right)^{1/4} \e^{- \frac{kp}{2} (x^{2m-1})^2} \sum_{l \in \mathbb{Z}} \e^{\im\pi \left(l + \frac{i_m}{pq_m} \right)^2 pq_m\tau_m} \e^{\left(l + \frac{i_m}{pq_m} \right) \frac{pq_m}{l_{2m}}(x^{2m-1} + \im x^{2m})}.
}
Here, we fixed the constant $c^{(p)}_{i_m} = \left( \frac{kp}{4 \pi^3 l_{2m}^2}\right)^{1/4}$.
Note that the index $i_m \in \{0,1,\cdots,pq_m-1\}$ is rather considered to be an element of the additive group $\mathbb{Z}/pq_m \mathbb{Z}$ because of the cyclic structure $f^{(p)}_{i_m}= f^{(p)}_{i_m + pq_m}$.
This basis is not only complete but also orthonormal.
In Appendix \ref{integration formula 2}, we show the relation,
\eq{
\label{orthonormality of f_i_m}
\int_{0}^{2\pi l_{2m-1}} \dif x^{2m-1}\int_{0}^{2\pi l_{2m}} \dif x^{2m} (f^{(p)}_{i_m})^* f^{(p)}_{j_m} = \delta_{i_m,j_m},}
which implies the orthonormality
\eq{ \int_{T^{2n}} \mu (f^{(p)}_{i})^* f^{(p)}_{j} = \prod_{m=1}^n \delta_{i_m,j_m}.}

Now, let us check that the number of zero modes is consistent with the index theorem and the vanishing theorem.
As we obtained in (\ref{torus zero mode 2}), the number of linearly independent zero modes is
\eq{ \dim \mathrm{Ker} \, D^{(p)} = p^n \prod_{m=1}^n q_m.}
On the other hand, the index theorem and the vanishing theorem implies
\eq{ \dim \mathrm{Ker} \, D^{(p)} = \mathrm{Ind} \, D^{(p)} = \int_{T^{2n}} \e^{\frac{\im p}{2\pi} R^L} = \frac{(kp)^n}{(2\pi)^n}\int_{T^{2n}} \mu = p^n \prod_{m=1}^n q_m.}

\subsection{Matrix regularization of embedding functions}
Now, let us consider the following functions:
\eq{ \label{u_n v_m}
u_m = \e^{\im x^{2m-1}/l_{2m-1}}, \quad v_m = \e^{\im x^{2m}/l_{2m}}.}
By using these functions, an isometric embedding $X^A: T^{2n} \to \mathbb{R}^{4n}$ can be written as
\als{X^{4m-3} &= \frac{l_{2m-1}}{2}(u_m + u_m^*), \quad X^{4m-2} = \frac{l_{2m-1}}{2\im}(u_m - u_m^*), \\
X^{4m-1} &= \frac{l_{2m}}{2}(v_m + v_m^*), \hspace{9mm} X^{4m} = \frac{l_{2m}}{2\im}(v_m - v_m^*).}
We consider the matrix regularization of these functions.

We define a matrix regularization of $C^{\infty}(T^{2n})$ by
\eq{ T_p (f) = \Pi^{(p)} f \Pi^{(p)}, \quad (f \in C^{\infty}(T^{2n})) }
where $\Pi^{(p)}: \Gamma(S \otimes L^{\otimes p}) \to \mathrm{Ker} \, D^{(p)}$ is the Hermitian projection.
Using the integral in Appendix \ref{integration formula 2}, (or see \cite{Adachi:2020asg}), we have
\als{
U^{(p)}_m := T_p(u_m) &= I_{pq_1} \otimes \cdots \otimes I_{pq_{m-1}} \otimes U_{pq_m} \otimes I_{pq_{m+1}} \otimes \cdots \otimes I_{pq_{n}},\\
V^{(p)}_m := T_p(v_m) &= I_{pq_1} \otimes \cdots \otimes I_{pq_{m-1}} \otimes V_{pq_m} \otimes I_{pq_{m+1}} \otimes \cdots \otimes I_{pq_{n}},
}
where
\als{
\label{U,V}
U_{pq_m} &= \e^{-\frac{1}{4 kp l_{2m-1}^2}} \left(
\begin{array}{cccc}
     1 &  & & \\
     & \e^{\im \frac{2\pi}{pq_m}} & & \\
     &  & \ddots&  \\
    &  & & \e^{\im \frac{2(pq_m-1)\pi}{pq_m}}
\end{array}
\right),\\
V_{pq_m} &=  \e^{-\frac{1}{4 kp l_{2m}^2}} \left(
\begin{array}{ccccc}
     &  & & &1\\
     1&  & & &\\
     & 1 & &  &\\
     &  &\ddots & &\\
    &  & & 1& 
\end{array}
\right).
}
These matrices satisfy the algebra of non-commutative torus $U_{pq_m}V_{pq_m}=\e^{\im \frac{2\pi}{pq_m}}V_{pq_m}U_{pq_m}$.
Therefore, the matrix regularization of the embedding functions are
\als{
\label{torus T(X^A)}
T_p(X^{4m-3}) &= \frac{l_{2m-1}}{2}(U^{(p)}_m + U^{(p)\dagger}_m), \quad T_p(X^{4m-2}) = \frac{l_{2m-1}}{2\im}(U^{(p)}_m - U^{(p)\dagger}_m), \\
T_p(X^{4m-1}) &= \frac{l_{2m}}{2}(V^{(p)}_m + V^{(p)\dagger}_m), \hspace{8mm} T_p(X^{4m}) = \frac{l_{2m}}{2\im}(V^{(p)}_m - V^{(p)\dagger}_m).}

\subsection{Laplace operator on $\Gamma(L^{\otimes q})$}

Consider the Laplace operator on $\Gamma(L^{\otimes q})$
\eq{ \Delta^{(q)} = - \sum_{a=1}^{2m} (D^{(q)}_a)^2 = - \sum_{m=1}^n \left( D^{(q)}_{m} D^{(q)}_{\bar{m}} + D^{(q)}_{\bar{m}}D^{(q)}_{m} \right),}
where $D^{(q)}_{a}$ is the connection of $\Gamma(L^{\otimes q})$ in the real coordinates $x^a$ and $D^{(q)}_{m}$ and $D^{(q)}_{\bar{m}}$ are those in the complex coordinates.
Also let us define the inner product
\eq{(f^{(q)},g^{(q)}):= \int_{T^{2n}} \mu (f^{(q)})^* g^{(q)}. \quad (f^{(q)},g^{(q)} \in \Gamma(L^{\otimes q}))}
Here, $\mu = \omega^{\wedge n}/n! = \dif x^1 \wedge \dif x^2 \wedge \cdots \wedge \dif x^{2n}$.

First, let us examine the spectrum of the Laplacian $\Delta^{(q)}$ for $q=0$, i.e. the case for the ordinary functions $C^{\infty}(T^{2n})$.
One can easily see that the normalized eigenfunctions of $\Delta^{(0)}$ are
\eq{ \label{f_b}
f_{b} (x) = [(2\pi)^{2n} l_1 l_2 \cdots l_{2n}]^{-1/2} \prod_{a=1}^{2n} \e^{\im b_a x^a/l_a},
}
and the eigenvalues are given by
\eq{ E_b = \sum_{a=1}^{2n} \left( \frac{b_a}{l_a} \right)^2 ,}
where $b=(b_1,b_2,\cdots, b_{2n}) \in \mathbb{Z}^{2n}$.
They satisfy
\eq{
\label{orthonormality of f_b}
(f_{b},f_{b'}) = \prod_{a=1}^{2n} \delta_{b_a,b'_a}.
}

Now, let us consider the spectrum of the Laplacian $\Delta^{(q)}$ for $q\neq 0$.
Since we have
\eq{ [D^{(q)}_{m}, D^{(q)}_{\bar{m}'}] = kq\delta_{m,m'}, \quad [D^{(q)}_{m}, D^{(q)}_{m'}] = [D^{(q)}_{\bar m}, D^{(q)}_{\bar{m}'}]=0,}
we can define creation and annihilation operators 
\eq{
\label{raising and lowering operators}
a^{(q)}_m := \im \frac{D^{(q)}_{\bar m}}{\sqrt{kq}}, \quad a^{(q)\dagger}_m := \im \frac{D^{(q)}_{m}}{\sqrt{kq}},
}
satisfying $[a^{(q)}_m, a^{(q)\dagger}_{m'}] = \delta_{m,m'}$.
Then, the Laplace operator can be written as
\eq{ \Delta^{(q)} = 2kq \sum_{m=1}^n \left( N^{(q)}_{m} + \frac{1}{2}\right),}
where $N^{(q)}_{m} := a^{(q)\dagger}_m a^{(q)}_m$ are the number operators.
Note that the lowest eigen modes of $\Delta^{(q)}$ should vanish under the action of $a^{(q)}_m \propto D^{(q)}_{\bar m}$ for all $m$.
This means that the lowest eigen modes are $f^{(q)}_{j}$ given in (\ref{torus zero mode 3}), which appeared in the discussion of the Dirac zero modes.
The other eigen modes are obtained by acting the creation operators on the lowest eigen modes $f^{(q)}_{j}$.
Thus, the normalized eigenfunctions of $\Delta^{(q)}$ are
\eq{
\label{Laplacian eigenvector T^2n}
f^{(q)}_{c,j} = \prod_{m=1}^n f^{(q)}_{c_m,j_m}, \quad f^{(q)}_{c_m,j_m} = \frac{(a^{(q)\dagger}_m)^{c_m}}{\sqrt{c_m!}} f^{(q)}_{j_m}(z^m),
}
and the corresponding eigenvalues are
\eq{ E_c = 2kq \sum_{m=1}^n \left( c_m + \frac{1}{2}\right).}
Here $c = (c_1,c_2,\cdots,c_n) \in  (\mathbb{Z}_{\ge 0})^n$.
More explicitly, the eigenfunctions are given by
\als{ \label{f_c,j}
f^{(q)}_{c_m,j_m}(z^m) =& \left( \frac{kq}{4 \pi^3 l_{2m}^2}\right)^{1/4} \frac{(-\im)^{c_m}}{\sqrt{2^{c_m} c_m !}} \e^{-\frac{kq}{2} (x^{2m-1})^2}\\
&\times \sum_{l \in \mathbb{Z}} \e^{\im\pi \left(l + \frac{j_m}{qq_m} \right)^2 qq_m\tau_m} \e^{\left(l + \frac{j_m}{qq_m} \right) \frac{qq_m}{l_{2m}}(x^{2m-1} + \im x^{2m})} H_{c_m}(\sqrt{kq} (x^{2m-1} - 2\pi l_{2m-1} (l+ \frac{j_m}{qq_m}))).}
Here, $H_n(x)$ is the Hermite polynomial satisfying the recursion $H_{n+1}(x) = 2x H_n(x) - H'_n(x)$.

\subsection{Matrix regularization of $\Gamma(L^{\otimes q})$ and the Laplace operator}

In this subsection, we explicitly evaluate the Toeplitz operator for a complete basis of $\Gamma(L^{\otimes q})$ given by the eigenfunctions of $\Delta^{(q)}$ and discuss the matrix Laplacian.

The matrix regularization of $\Gamma(L^{\otimes q})$ is defined by
\eq{ T_p (f^{(q)}) = \Pi^{(p+q)} f^{(q)} \Pi^{(p)}, \quad (f^{(q)} \in \Gamma(L^{\otimes q})) }
where $\Pi^{(p)}: \Gamma(S \otimes L^{\otimes p}) \to \mathrm{Ker} \, D^{(p)}$ is the Hermitian projection.

For $q =0$, we have the eigenfunctions $f_b$ given in (\ref{f_b}).
Using the results of Appendix \ref{integration formula 2}, we have
\als{  T_p (f_{b}) &= [(2\pi)^{2n} l_1 l_2 \cdots l_{2n}]^{-1/2} \e^{- \frac{1}{4kp} \sum_{m=1}^n\left( \frac{b_{2m-1}^2-b_{2m-1}}{l_{2m-1}^2} + \im \frac{2b_{2m-1}b_{2m}}{l_{2m-1} l_{2m}}+  \frac{b_{2m}^2-b_{2m}}{l_{2m}^2} \right)}\\
& \quad \times (U^{(pq_1)})^{b_{1}} (V^{(pq_1)})^{b_{2}} \otimes (U^{(pq_2)})^{b_{3}} (V^{(pq_2)})^{b_{4}} \otimes \cdots \otimes (U^{(pq_n)})^{b_{2n-1}} (V^{(pq_n)})^{b_{2n}}.}

For $q \neq 0$, the Toeplitz operators of the eigenfunctions $f^{(q)}_{c,j}$ given in (\ref{f_c,j}) are
\eq{
\label{torus Toeplitz operator of eigenfunctions}
T_p (f^{(q)}_{c,j})_{i,i'} =  \prod_{m=1}^n \int_{0}^{2\pi l_{2m-1}} \dif x^{2m-1}  \int_{0}^{2\pi l_{2m}} \dif x^{2m} (f^{(p+q)}_{i_m})^* f^{(q)}_{c_m,j_m} f^{(p)}_{i'_m},
}
where $i=(i_1,i_2,\cdots i_n)$ and $i' = (i'_1,i'_2,\cdots i'_n)$ are the labels of the Dirac zero modes.
The integral of the right-hand side of (\ref{torus Toeplitz operator of eigenfunctions}) is computed in Appendix \ref{eigenstate products} and the result is
\als{ \label{fuzzy torus T_p(f)}
&\int_{0}^{2\pi l_{2m-1}} \dif x^{2m-1}  \int_{0}^{2\pi l_{2m}} \dif x^{2m} (f^{(p+q)}_{i_m})^* f^{(q)}_{c_m,j_m} f^{(p)}_{i'_m}\\
&= \frac{\im^{c_m}}{\sqrt{2^{c_m} c_m !}} \left( \frac{kq}{4 \pi^3 l_{2m}^2}\right)^{1/4} \left( \frac{p}{p+q }\right)^{\frac{c_m}{2} + \frac{1}{4}} \sum_{t=1}^{(p+q)q_m} \delta^{(\mathrm{mod} \ (p+q)q_m)}_{i_m, j_m + i'_m + qq_m t} \sum_{l \in \mathbb{Z}} \\
& \times \e^{\im\pi \left(l + \frac{p q_m i_m - (p+q) q_m i'_m}{(p+q)pqq_m^3} \right)^2 (p+q)pqq_m^3 \tau_m} H_{c_m}(2\pi l_{2m-1} \sqrt{k(p+q)pqq_m^2} (l+ \frac{p q_m i_m - (p+q) q_m i'_m}{(p+q)pqq_m^3})).}

From (\ref{fuzzy matrix laplacian}), we define a Laplace operator acting on $T_p (f^{(q)})$ by
\als{ \hat{\Delta}^{(q)} T_p(f^{(q)}) 
&= (kp)^{2} [T(X^A),[T(X^A),T_p(f^{(q)})]] \\
&= \frac{l_{2m-1}^2 (kp)^2}{2} \sum_{m=1}^n \left(  [T(u_m),[T(u_m)^{\dagger},T_p(f^{(q)})]] + [T(u_m)^{\dagger}, [T(u_m), T_p(f^{(q)})]] \right)\\
& \quad + \frac{l_{2m}^2 (kp)^2}{2} \sum_{m=1}^n \left(  [T(v_m),[T(v_m)^{\dagger},T_p(f^{(q)})]] + [T(v_m)^{\dagger}, [T(v_m), T_p(f^{(q)})]] \right).}
The second expression is obtained by using (\ref{torus T(X^A)}).
For $q=0$, we can easily see that the spectrum of $\hat{\Delta}^{(0)}$ approaches that of $\Delta^{(0)}$ as
\als{ \hat{\Delta}^{(0)} T_p(f_b) 
&= 4(kp)^2 \sum_{m=1}^n \left( l_{2m-1}^2 \sin^2(\frac{\pi b_{2m}}{pq_m}) + l_{2m}^2 \sin^2(\frac{\pi b_{2m-1}}{pq_m}) \right) T_p(f_b) \\
&= \left(\sum_{a=1}^{2n} \left( \frac{b_a}{l_a} \right)^2 + O(p^{-1})\right)T_p(f_b).}
We can also see the correspondence between the trace and the integral.
In fact, we have
\begin{align}
(2\pi \hbar_p)^n \Tr [T_p(f_b)^{\dagger}T_p(f_{b'})] &= (2\pi \hbar_p)^n [(2\pi)^{2n} l_1 l_2 \cdots l_{2n}]^{-1} \e^{- \frac{1}{2kp} \sum_{m=1}^n\left( \frac{b_{2m-1}^2-b_{2m-1}}{l_{2m-1}^2} + \im \frac{2b_{2m-1}b_{2m}}{l_{2m-1} l_{2m}}+  \frac{b_{2m}^2-b_{2m}}{l_{2m}^2} \right)} \nonumber \\
& \quad \times \prod_{m=1}^{n}(\e^{-\frac{b_{2m-1}}{2 kp l_{2m-1}^2}} \e^{-\frac{b_{2m}}{2 kp l_{2m}^2}} pq_m) \delta^{(\mathrm{mod} \, pq_m)}_{b_m,b'_m} \nonumber \\
&= \delta^{(\mathrm{mod} \, pq_m)}_{b_m,b'_m} +O(p^{-1}),
\end{align}
which is consistent with (\ref{orthonormality of f_b}).
For $q \neq 0$, the eigenvalue problem of the Laplace operator $\hat{\Delta}^{(q)}$ is related to the Hofstadter problem as noted in \cite{Adachi:2020asg}.
It is numerically shown in \cite{Adachi:2020asg} that the spectrum of $\hat{\Delta}^{(q)}$ approaches that of $\Delta^{(q)}$ in the commutative limit
\footnote{In \cite{Adachi:2020asg}, only the two-dimensional case is considered, while we consider higher dimensional torus $T^{2n}$.
However, $T^{2n}$ can be decomposed to the tensor product of $T^{2}$s so that the results of \cite{Adachi:2020asg} can also be applied to our case.}.

\section{Conclusion and future problems}

In this paper, we studied the Berezin-Toeplitz quantization of vector bundles over a general closed connected K\"ahler manifold, which is a continuation of our previous studies of two-dimensional cases \cite{Adachi:2021ljw, Adachi:2021aux}.
In our formalism, we treated a vector bundle as a homomorphism bundle and treat its sections as some linear operator between suitable twisted spinor fields.
By restricting the vector spaces of each twisted spinor fields to finite-dimensional kernels of Dirac operators, we defined a quantization map from fields (sections of the vector bundle) to matrices.
We obtained a large-$p$ asymptotic behavior of the product $T_p(\varphi) T_p(\chi)$ for arbitrary sections of vector bundles $\varphi,\chi$, up to the second order in $1/p$.
This is a natural generalization of the relation of matrix regularization (\ref{MatReg}).
The matrix Laplacian acting of such matrices can be written in terms of a commutator-like operation and its spectrum in the large-$p$ limit is shown to be equal to that of the usual Bochner Laplacian acting on continuum fields.
Our result is a generalization of \cite{Carow-Watamura:2004mug,Dolan:2006tx}, where fuzzy $CP^n$ is considered, to the general K\"ahler manifold.
As explicit examples, we considered monopole bundles over a fuzzy $CP^n$ and fuzzy $T^{2n}$ and we confirmed that in the case of $CP^n$ our formulation correctly reproduces the results in \cite{Carow-Watamura:2004mug,Dolan:2006tx}.

Our framework is applicable to a wide class of fields.
For example, an $(r,s)$ tensor field gives a homomorphism from $\Gamma(TM^{\otimes s})$ to $\Gamma(TM^{\otimes r})$, and we can apply our formulation.
It is interesting to construct a fuzzy version of the higher spin theories \cite{Fronsdal:1978rb,Vasiliev:1999ba} by using our method.
It is also possible to consider a matrix regularization of spinor fields.
The spinor fields on the lattice have the problems of doublers and chiral anomaly and we can consider similar problems on fuzzy spaces \cite{Grosse:1994ed, Carow-Watamura:1996uej,Balachandran:1999qu,Aoki:2002fq,Aoki:2002mn,Balachandran:2003ay}.
Our method will enable us to deal with the similar problems on a general K\"ahler manifold.
Our method can also be used to construct fuzzy field theories in arbitrary background fields.
It is important to understand how various background field configurations such as the instantons are realized on fuzzy spaces.

Let us comment on some possible generalizations of our study.
Throughout this paper, we assumed that the manifold $M$ is K\"ahler.
In particular, we assumed that the manifold has the integrable complex structure.
However, it is possible to construct a quantization of functions with almost complex structure which is not necessarily integrable (for example see \cite{Ma-Marinescu}).
Moreover, it is also possible to consider non-compact manifolds and orbifolds \cite{Ma-Marinescu}.
Therefore, the Berezin-Toeplitz quantization of vector bundles might also be defined over more general manifolds than the closed K\"ahler case (for example the fuzzy $S^4$ \cite{Medina:2002pc,Abe:2004sa,Hasebe:2020nmt,Hasebe:2021vvd}).
We can also consider more challenging problems such as a quantization of odd-dimensional manifolds \cite{Awata:1999dz,Dolan:2003kq,Yoneya:2016wqw,Hasebe:2017myo} or manifolds with boundaries.
These studies are important to uncover how the various branes of such geometries such as odd-dimensional branes and orientifold planes \cite{Itoyama:1998et} in the framework of matrix models \cite{Yoneya:2016wqw}.
Yet another possible generalization is the Berezin-Toeplitz quantization of nonlocal operators such as Wilson lines.
As a Wilson line send a spinor at a point to a spinor at a different point, it gives a linear map between twisted spinor spaces.
Wilson line or loop is an essential ingredient of gauge theories and the quantization of Wilson lines may shed light on studies of gauge theories on fuzzy geometries.

Finally, another direction of the study of fuzzy spaces is the inverse problem of quantization.
While, in quantization, one constructs a quantum geometry from a given classical geometry, it is also interesting to consider the problem of finding a classical geometry from a given quantum geometry.
See \cite{Shimada:2003ks,Berenstein:2012ts,Ishiki:2015saa,Schneiderbauer:2016wub,Ishiki:2016yjp,Asakawa:2018gxf,Ishiki:2018aja,Terashima:2018tyi,Sako:2022pid} for developments in this direction.
We consider that the inverse problem can be generalized for the case of vector bundles.
The matrix counterparts of vector bundles should contain various geometric information and finding a method of extracting such information will bring great progress for understanding the fuzzy geometry.

\section*{Acknowledgments}

The work of H. A. and G. I. was supported  
by JSPS KAKENHI (Grant Numbers 21J12131 and 19K03818, respectively). 
This work of S. K. was supported by JST, the establishment of university fellowships towards the creation of science technology innovation, Grant Number JPMJFS2106.

\appendix
\numberwithin{equation}{section}
\setcounter{equation}{0}

\section{Proofs and formulas for general K\"ahler manifolds}

\subsection{Useful choice of orthonormal frame fields}
\label{complex orthonormal}
In this Appendix, we will introduce particular orthonormal frame fields (vielbeins) which simplifies our argument.

Let us choose an element $e_1\in \Gamma(TM)$ such that $g(e_1,e_1) =1$.
Then, $e_2 := J e_1 \in \Gamma(TM)$ satisfies $g(e_a,e_b) =\delta_{ab}$ for $a,b=1,2$, which follows from the K{\"a}hler condition (\ref{Kahler}).
Next, choose an arbitrary $e_3\in \Gamma(TM)$ such that $g(e_a,e_b) =\delta_{ab}$ for $a,b=1,2,3$.
Then, $e_4 := J e_3 \in \Gamma(TM)$ also satisfies $g(e_a,e_b) = \delta_{ab}$ for $a,b=1,2,3,4$.
By continuing the above argument, we can construct a complete orthonormal fields.
This choice is useful because the symplectic form can be written be
\eq{
\label{symplectic 1}
\omega = \sum_{m=1}^n \theta^{2m-1} \wedge \theta^{2m}
}
where $\{\theta^a\}_{a=1,2,\cdots,2n}$ is the dual basis of $\{e_a\}_{a=1,2,\cdots,2n}$.

It is also convenient to introduce complexified fields
\eq{
    \label{complex vielbein}
    w_m := \frac{1}{\sqrt 2} (e_{2m-1} - \im  e_{2m}), \quad \bar{w}_{m} := \frac{1}{\sqrt 2} (e_{2m-1} + \im  e_{2m}),
}
for $m=1,2,\cdots,n$.
Note that the properties, $J w_m = \im w_m$ and $J \bar{w}_{m} = - \im \bar{w}_{m}$, imply $w_m$ and $\bar{w}_m$ are holomorphic and antiholomorphic vector fields, respectively.
In this frame, the metric components are
\eq{ g(w_m,\bar{w}_{l})  = g(\bar{w}_{m}, w_l) = \delta_{ml}, \quad g(w_m,w_l) =  g(\bar{w}_{m},\bar{w}_{l})=0.}

\subsection{Gamma matrices in Weyl representation}
\label{Clifford}
In this Appendix, we will consider the gamma matrices in the Weyl representation for a $2n$-dimensional manifold.

Let $\{\gamma^a_{(2n)}\}_{a=1,2,\cdots,2n}$ be a set of square matrices with size $2^n$.
They are called the gamma matrices when they satisfy the Clifford algebra for $\mathbb{R}^{2n}$,
\eq{
    \{\gamma^a_{(2n)}, \gamma^b_{(2n)}\} = 2 \delta^{ab} I_{2^n}.
}
Here, $\{ \ , \ \}$ is the anti-commutator and $I_{2^n}$ is the identity matrix with size $2^n$.
We can also define a chirality matrix by
\eq{
\gamma_{(2n)} := (-\im)^n \gamma^1_{(2n)}\gamma^2_{(2n)} \cdots \gamma^{2n}_{(2n)}.
}
This matrix is Hermitian and anticommutes with all of the gamma matrices $\{\gamma_{(2n)},\gamma^a_{(2n)}\} = 0$.
We can choose a representation such that $\gamma_{(2n)} = \sigma^3 \otimes I_{2^{n-1}}$, where $\{\sigma^a\}_{a=1,2,3}$ are the Pauli matrices and $\otimes$ is the Kronecker product.
The Weyl (chiral) representation can be constructed by the recursion
\als{
\label{recursion}
&\gamma_{(2)}^1 = \sigma^1, \ \gamma_{(2)}^2 = \sigma^2, \\
&\gamma_{(2n+2)}^i =\sigma^2  \otimes \gamma_{(2n)}^i \quad (i=1,2,\cdots,2n),\\
&\gamma_{(2n+2)}^{2n+1} = \sigma^2 \otimes \gamma_{(2n)}, \\
&\gamma_{(2n+2)}^{2n+2} = - \sigma^1 \otimes I_{2^{n}}.
}
We extensively use these relations in proofs given in the following appendices.

Also, consider the gamma matrices in the complex orthonormal frame defined by
\eq{\gamma_{(2n)}^m := \frac{\gamma_{(2n)}^{2m-1} + \im \gamma_{(2n)}^{2m}}{\sqrt{2}}, \quad \gamma_{(2n)}^{\bar m} := \frac{\gamma_{(2n)}^{2m-1} - \im  \gamma_{(2n)}^{2m}}{\sqrt{2}}.}
These matrices satisfy $\{\gamma_{(2n)}^m,\gamma_{(2n)}^{\bar l} \} = 2\delta_{ml} I_{2^n}, \ \{\gamma_{(2n)}^m,\gamma_{(2n)}^l \} = \{\gamma_{(2n)}^{\bar m},\gamma_{(2n)}^{\bar l} \} = 0$ and $(\gamma_{(2n)}^m)^{\dagger} = \gamma_{(2n)}^{\bar m}$.
Let $\ket{\pm}$ be the normalized eigenvector of $\sigma^3$ with eigenvalue $\pm1$.
Then, we can recursively show the important properties,
\als{ \label{gamma matrices identities}
&c_m \gamma_{(2n)}^{\bar{m}}\ket{+}^{\otimes n} = 0 \quad \Rightarrow \quad c_m = 0,\\
&\gamma_{(2n)}^{m} \ket{+}^{\otimes n} = 0, \quad \gamma_{(2n)}^m \gamma_{(2n)}^{\bar l} \ket{+}^{\otimes n} = 2\delta_{ml}\ket{+}^{\otimes n},
}
where $c_m$ is a complex number.

\subsection{Vanishing theorem and index theorem}
\label{Vanishing}

Let $D_i$ be the Dirac operator on $\Gamma(S_c \otimes L^{\otimes p} \otimes E_i)$.
In this Appendix, we will show that the zero modes of $D_i$ have the positive chirality and ${\rm dim}\mathrm{Ker}\, D_i = \mathrm{rank}(E_i) (2\pi \hbar_p)^{-n} \int_M \mu + O(p^{n-1})$ for sufficiently large $p$.
The former is known as the vanishing theorem and the latter is a consequence of 
the index theorem.
We also show that nonzero eigenvalues of 
$D_i$ has a large gap of $O(\sqrt{p})$.
As a notational brevity, we will omit superscript of covariant derivative and simply write $\nabla$ and we also omit the identity operators unless required.

The chirality operator $\gamma_{(2n)} = I_{2^{n-1}} \otimes \sigma^3$ anticommutes with $D_i$ and we find 
\eq{
\label{D_i Weyl}
    D_i =
    \left(
    \begin{array}{cc}
       0 & D_i^{-} \\
    D_i^{+} & 0
    \end{array}
    \right).
}
Here, $\pm$ indicates the chirality of the space on which the operators are acting.

We first compute the square of $D_i$, which is needed to show $\mathrm{Ker} D_i^{-} = \{0\}$ for large enough $p$.
From (\ref{D_i Weyl}), we have
\eq{
    (D_i)^2 = 
    \left(
    \begin{array}{cc}
        D_i^{-} D_i^{+} & 0 \\
        0 & D_i^{+}D_i^{-}
    \end{array}
    \right).
    \label{d square 1}
}
We also use the Weitzenb$\ddot{\text{o}}$ck formula,
\eq{
    \label{D^2}
    (D_i)^2 = -\nabla_{a}\nabla_{a} +\frac{\im}{2}\hbar_p^{-1} \gamma^{a}_{(2n)}\gamma^{b}_{(2n)} \omega_{ab} - \frac{1}{2}\gamma^{a}_{(2n)}\gamma^{b}_{(2n)}R^{S_c \otimes E_i}_{ab}, 
}
where $R^{S_c \otimes E_i}_{ab}:= R^{S_c \otimes E_i}(e_a,e_b)$.
Let us introduce differential operators
\eq{
\nabla_m := \nabla_{w_m} = \frac{1}{\sqrt 2} (\nabla_{2m-1} - \im \nabla_{2m}),  \quad \nabla_{\bar m} := \nabla_{\bar{w}_m} = \frac{1}{\sqrt 2} (\nabla_{2m-1} + \im \nabla_{2m}).
}
Employing these operators, we have for fixed $m$,
\als{
\nabla_{2m-1}\nabla_{2m-1}+\nabla_{2m}\nabla_{2m} &= 2\nabla_m \nabla_{\bar m} - [\nabla_m,\nabla_{\bar m}] \\
&= 2\nabla_m \nabla_{\bar m} -\hbar_p^{-1} - R^{S_c \otimes E_i}_{m\bar{m}}.
}
Here, we used (\ref{symplectic 1}) and (\ref{complex vielbein}) in the last equality and $R^{S_c \otimes E_i}_{m \bar{m}} := R^{S_c \otimes E_i}(w_m,\bar{w}_m)$.
Using the above equation, the first term of (\ref{D^2}) can be written as
\eq{
-\nabla_{a}\nabla_{a}= -2 \nabla_m \nabla_{\bar m} +n \hbar_p^{-1} + R^{S_c \otimes E_i}_{m\bar{m}},
}
where the repeated indices $a$ and $m$ are summed.
Hence, we have
\eq{
\label{D^2 n}
(D_i)^2 = -2 \nabla_m \nabla_{\bar m} + \hbar_p^{-1} A_n  + R_i,
}
where
\als{A_n &:= n + \frac{ \im }{2} \gamma^{a}_{(2n)}\gamma^{b}_{(2n)} \omega_{ab},\\
R_i &:=  - \frac{1}{2}\gamma^{a}_{(2n)}\gamma^{b}_{(2n)}R^{S_c \otimes E_i}_{ab} + R^{S_c \otimes E_i}_{m\bar{m}}.}
More explicitly, $R^{S_c \otimes E_i}$ is given by
\eq{ R^{S_c \otimes E_i} = R^S + \frac{1}{2}R^{L_c} + R^{E_i}, \quad  R^S_{ab} = \frac{1}{4} R_{abcd} \gamma_{(2n)}^c \gamma_{(2n)}^d, \quad  R^{L_c}_{ab} = - R_{ab m \bar{m}},}
where $R_{abcd}$ is the Riemann curvature tensor.
Then, we have
\eq{ \label{R_i}
R_i = \frac{1}{2} R + \frac{1}{2} \gamma_{(2n)}^a \gamma_{(2n)}^b R_{abm\bar{m}} - \frac{1}{2}\gamma^{a}_{(2n)}\gamma^{b}_{(2n)}R^{E_i}_{ab} + R^{E_i}_{m\bar{m}},}
where $R$ is the scalar curvature and we used $\gamma_{(2n)}^a \gamma_{(2n)}^b \gamma_{(2n)}^c \gamma_{(2n)}^d R_{abcd} = - 2 R$ and $R_{m \bar{m} l \bar{l}} = - \frac{1}{2}R$.

$2^{n}\times 2^n$ matrix $A_n$ has following properties if we use the Weyl representation discussed in Appendix \ref{Clifford}.
First property is that $A_n$ is diagonal and positive semidefinite.
This can be shown recursively as follows.
From (\ref{recursion}), one obtains $A_{n+1} = I_2 \otimes A_n + I_{2^{n+1}} - \sigma^3 \otimes \gamma_{(2n)}$ and it shows that if $A_n$ is diagonal and positive semidefinite, so is $A_{n+1}$.
By checking $A_1= I_2-\sigma^3$, which is obviously diagonal and positive semidefinite, we proved the first property.
Second property of $A_n$ is that its eigenvector with eigenvalue $0$ is proportional to $\ket{+}^{\otimes n}$.
This can be shown by the similar recursive method.

Let us use (\ref{D^2 n}) to prove $\mathrm{Ker} D_i^{-} = \{0\}$ for large enough $p$.
For any $\psi\in \Gamma(S_c \otimes L^{\otimes p} \otimes E_i)\setminus \{0\}$, we have
\eq{
\label{norm of D^2}
|D_i \psi|^2 = 2 | \nabla_{\bar m} \psi|^2 + \hbar_p^{-1} (\psi,A_n \psi) + (\psi, R_i \psi) \ge \hbar_p^{-1} (\psi,A_n \psi) - |R_i| |\psi|^2.
}
For $\psi$ which is not proportional to $\ket{+}^{\otimes n}$, $(\psi,A_n \psi)$ is strictly positive.
Therefore, for sufficiently large $p$ satisfying $\hbar_p^{-1} > |R_i| |\psi|^2/(\psi,A_n \psi)$, the right-hand side of (\ref{norm of D^2}) becomes positive and it implies that $D_i \psi \neq 0$.
This means that the Dirac zero modes must be proportional to $\ket{+}^{\otimes n}$ for sufficiently large $p$.
Since $\ket{+}^{\otimes n}$ has the positive chirality, we conclude that 
$\mathrm{Ker} D_i^{-} = \{0\}$ for large enough $p$.

We next show that $\dim \mathrm{Ker}\, D_i = \mathrm{rank}(E_i) (2\pi \hbar_p)^{-n} \int_M \mu + O(p^{n-1})$.
Note that, when $\mathrm{Ker} D_i^{-} = \{0\}$, 
we have the following relations:
\eq{
\mathrm{dim} \, \mathrm{Ker} \, D_i = \mathrm{dim} \, \mathrm{Ker} \, D_i^{+} = \mathrm{Ind} \, D_i.
}
On the other hand, the Atiyah-Singer index theorem states that
\eq{
    \label{index}
    \mathrm{Ind}D_i = \int_M \mathrm{Td}(T^{(1,0)}M) \wedge \mathrm{ch}(L^{\otimes p} \otimes E_i)
}
Here, $\mathrm{Td}(\cdot)$ and $\mathrm{ch}(\cdot)$ are the Todd class and the Chern character, respectively, and $T^{(1,0)}M$ is the holomorphic tangent bundle.
By expanding in $p$, we find
\eq{
\mathrm{dim}\mathrm{Ker}\, D_i  = \mathrm{rank}(E_i) \int_M \e^{ \frac{\im p}{2\pi} R^L} + O(p^{n-1}) = \frac{\mathrm{rank}(E_i)}{(2\pi \hbar_p)^n} \int_M \mu + O(p^{n-1}) .
}

Finally, we prove that nonzero eigenvalues of $D_i$ have a large gap 
of $O(\sqrt{p})$.
Let $\lambda$ be a non-zero eigenvalue of $D_i$.
Then, the eigenvalue equation for $(D_i)^2$ is equivalent to
\eq{
    \begin{cases}
        D_i^{-} D_i^{+} \psi^+ &= \lambda^2 \psi^+,\\
        D_i^{+} D_i^{-}  \psi^- &= \lambda^2 \psi^-,
    \end{cases}
}
for $\psi \in \Gamma(S_c \otimes L^{\otimes p} \otimes E_i) \setminus \{0\}$, where $\psi^{\pm}$ is the positive/negative chirality mode of $\psi$.
If $\psi^- \neq 0$, (\ref{norm of D^2}) implies that 
$\lambda^2 \ge O(p)$.
If $\psi^- = 0$, we have $\psi^+ \neq 0$ in order for $\psi$ to be nonzero.
By using the relation $D_i^{+} D_i^{-} (D_i^{+} \psi^+) = \lambda^2 (D_i^{+} \psi^+$), we again find that (\ref{norm of D^2}) implies $\lambda^2 \ge O(p)$. Thus, in any case, we have $\lambda^2 \ge O(p)$.
This shows that $\lambda^2$ is at least of $O(p)$ and thus, the nonzero eigenvalues 
of $D_i$ indeed have a gap of at least $O(\sqrt{p})$.

\subsection{Asymptotic expansion for Toeplitz operators}
\label{asymptotic expansion}
In this Appendix, we compute the product $T_p^{(E_1,E_2)}(\varphi)T_p^{(E_2,E_3)}(\chi)$ for $\varphi \in \Gamma(\mathrm{Hom}(E_2,E_1))$ and $\chi \in \Gamma(\mathrm{Hom}(E_3,E_2))$
and show that it can be expanded in a power series of $\hbar_p$ for sufficiently large $p$.
The computation technique used in this Appendix is based on \cite{Hawkins:2005}.

First, we compute
\als{
    \label{asymptotic1}
    T_p^{(E_1,E_2)}(\varphi)T_p^{(E_2,E_3)}(\chi) &= \Pi_1 \varphi \Pi_2 \chi \Pi_3 \\
    &= T_p^{(E_1,E_3)}(\varphi \chi) - \Pi_1 \varphi (1-\Pi_2) \chi \Pi_3.
}
For the computation of $1-\Pi_2$, let us consider the following Hermitian operator on $\Gamma(S_c \otimes L^{\otimes p} \otimes E_2)$:
\eq{
    P_2 := 
    \left(
    \begin{array}{cc}
        0 & D_2^{-} (D_2^{+} D_2^{-})^{-1} \\
        (D_2^{+} D_2^{-})^{-1} D_2^{+} & 0
    \end{array}
    \right).
}
Note that, since $\mathrm{Ker}D_2^{-} = \mathrm{Ker}D_2^{+} D_2^{-} = \{0\}$ for sufficiently large $p$ as shown in Appendix~\ref{Vanishing}, the inverse of
$D_2^{+} D_2^{-}$ always exists.
Let us consider the following combination,
\eq{
\label{D-P identity}
D_2 P_2 = P_2 D_2 = 
    \left(
    \begin{array}{cc}
         D_2^{-} (D_2^{+} D_2^{-})^{-1} D_2^{+}  & 0\\
        0 & 1
    \end{array}
    \right).
}
This gives the projection onto $(\mathrm{Ker}D_2)^{\perp}$, which should be equivalent to $1-\Pi_2$.
Thus, we find that
\eq{
    \label{projection}
    1-\Pi_2 = D_2 P_2 = D_2 (P_2)^2 D_2.
}
By using (\ref{asymptotic1}) and (\ref{projection}), for $\psi \in \mathrm{Ker}D_1$ and $\phi \in \mathrm{Ker}D_3$, we obtain
\als{
    \label{asymptotic2}
    (\psi,T_p^{(E_1,E_2)}(\varphi) T_p^{(E_2,E_3)}(\chi) \phi) &= (\psi,T_p^{(E_1,E_3)}(\varphi \chi) \phi) - (\psi,\varphi D_2 (P_2)^2 D_2 \chi \phi)\\
    &= (\psi,T_p^{(E_1,E_3)}(\varphi \chi) \phi) + (\psi, \varphi'  (P_2)^2 \chi' \phi).
    }
Here, we introduced the notation $\varphi' := \im  \gamma_{(2n)}^a\nabla_a \varphi$.
We also used $D_1 \psi = D_3 \phi = 0$ and
\eq{
\nabla^{E_1}(\varphi \varphi_2) = (\nabla^{\mathrm{Hom}(E_2,E_1)} \varphi) \varphi_2 + \varphi (\nabla^{E_2} \varphi_2),
}
for $\varphi_2 \in \Gamma(E_2)$.
Because $\gamma_{(2n)}^b\phi$ has the chirality $-1$, $\chi' \phi$ is in $(\mathrm{Ker}D_2)^{\perp}$.
On $(\mathrm{Ker}D_2)^{\perp}$, the projection
$1-\Pi_2 = D_2 P_2$ is the identity operator, which implies that $P_2$ is the inverse of $D_2$.
Thus, (\ref{asymptotic2}) can be written as
\eq{
    \label{asymptotic3}
    (\psi,T_p^{(E_1,E_2)}(\varphi) T_p^{(E_2,E_3)}(\chi) \phi) = (\psi,T_p^{(E_1,E_3)}(\varphi \chi) \phi) + (\psi, \varphi' (D_2)^{-2} \chi' \phi).
}
Let us then calculate $(D_2)^{-2}$ acting on $\chi'\phi$.
By using $A_n \gamma_{(2n)}^b \ket{+}^{\otimes n} = 2 \gamma_{(2n)}^b \ket{+}^{\otimes n}$, which can be obtained from (\ref{gamma matrices identities}), we have
\als{
    \label{asymptotic4}
    (D_2)^{-2} &= \left(-2 \nabla_m \nabla_{\bar m} + 2\hbar_p^{-1} + R_2\right)^{-1} \\
    &= \frac{\hbar_p}{2} - \frac{\hbar_p}{2} (D_2)^{-2} R_2 + \hbar_p (D_2)^{-2}  \nabla_m \nabla_{\bar m},
}
on $\chi'\phi$.
From $D_3 \phi=0$, one can obtain $\nabla_{\bar m} \phi =0$ (see also Appendix \ref{simplification}).
Then, (\ref{asymptotic3}) becomes
\eq{
    (\psi,T_p^{(E_1,E_2)}(\varphi) T_p^{(E_2,E_3)}(\chi) \phi) = (\psi,T_p^{(E_1,E_3)}(\varphi \chi) \phi) + \frac{\hbar_p}{2}(\psi, \varphi'\chi' \phi) + \epsilon,
}
where
\als{
    \epsilon &= \epsilon_1 + \epsilon_2,\\
    \epsilon_1 &= - \frac{\hbar_p}{2}(\psi, \varphi' (D_2)^{-2} R_2 \chi' \phi)\\
    \epsilon_2 &= \hbar_p (\psi, \varphi'(D_2)^{-2} (\nabla_m \nabla_{\bar m} \chi') \phi) + \hbar_p (\psi, \varphi'(D_2)^{-2} (\nabla_{\bar m} \chi') \nabla_m \phi).
}
Let us estimate the order of $\epsilon$ with respect to $\hbar_p$. 
If we set $\phi, \psi, \varphi$ and $\chi$ to $O(\hbar_p^0)$, the nontrivial $p$-dependences only appear in $\nabla_{m}\phi$ and $(D_2)^{-2}$.
As we discussed in Appendix~\ref{Vanishing}, all eigenvalues of $(D_2)^2$ 
are in the range $[C_1 \hbar_p^{-1}- C_2, \infty)$, where $C_1$ and $C_2$ are $p$-independent constants. Hence, the eigenvalues of $(D_2)^{-2}$ are in $(0,(C_1 \hbar_p^{-1} - C_2)^{-1}]$.
From this property and the fact that
the norm of a positive operator is equal to its maximum eigenvalues,
we find that $|(D_2)^{-2}| = O(\hbar_p)$.
For $\nabla_{m}\phi$, we can calculate
\als{
\label{norm order 1}
|\nabla_{m}\phi|^2 &= - (\phi,\nabla_{\bar m}\nabla_{m}\phi)
= (\phi,[\nabla_{m},\nabla_{\bar m}]\phi) - (\phi,\nabla_{m}\nabla_{\bar m}\phi)\\
&= \hbar_p^{-1} |\phi|^2 -  (\phi,R^{S_c \otimes E_3}_{m \bar{m}}\phi)  \\
&= O(\hbar_p^{-1}).
}
From these estimations, it follows that
\eq{
    |\epsilon_1|= O(\hbar_p^2), \quad  |\epsilon_2| = O(\hbar_p^{3/2}).
}
Then, we obtain
\eq{
    (\psi,T_p^{(E_1,E_2)}(\varphi)T_p^{(E_2,E_3)}(\chi) \phi) = (\psi,T_p^{(E_1,E_3)}(\varphi \chi) \phi) - \frac{\hbar_p}{2}(\psi, (\nabla_a \varphi)  (\nabla_b \chi) \gamma^a_{(2n)} \gamma^b_{(2n)}\phi) + O(\hbar_p^{3/2}).
}
From (\ref{gamma matrices identities}), we have
\eq{
\label{asymptotic C_0, C_1}
    (\psi,T_p^{(E_1,E_2)}(\varphi)T_p^{(E_2,E_3)}(\chi) \phi) = (\psi,T_p^{(E_1,E_3)}(\varphi \chi) \phi) - \hbar_p(\psi, (\nabla_m \varphi)  (\nabla_{\bar m} \chi) \phi) + O(\hbar_p^{3/2}).
}
From the asymptotic expansion of the Bergmann kernel \cite{asym}, the products of the Toeplitz operators also allow asymptotic expansion of integer power.
Thus, the $O(\hbar_p^{3/2})$ term is actually further bounded to $O(\hbar_p^{2})$.
The expansion (\ref{asymptotic C_0, C_1}) reproduces $C_0(\varphi,\chi)$ and $C_1(\varphi,\chi)$ in (\ref{asymptotic exp}).
This can be checked by noticing that the tensor $G^{ab} = g^{ab} + \im  W^{ab}$ has components $G^{m \bar{l}}=2 \delta_{ml}, \ G^{\bar{m}l} = G^{m l} = G^{\bar{m} \bar{l}} =0$.

One can also evaluate $C_2(\varphi,\chi)$ by recursively using (\ref{asymptotic4}).
Applying (\ref{asymptotic4}) to $\epsilon_1$, one finds
\als{ \epsilon_1 &= - \frac{\hbar_p}{4}(\psi, \varphi' \left( \hbar_p - \hbar_p  (D_2)^{-2} R_2  + 2\hbar_p (D_2)^{-2}  \nabla_m \nabla_{\bar m} \right) R_2 \chi' \phi)\\
&= - \frac{\hbar_p^2}{4}(\psi, \varphi' R_2 \chi' \phi) + O(\hbar_p^{5/2}).
}
Applying (\ref{asymptotic4}) to $\epsilon_2$, one finds
\als{ \epsilon_2 &= \frac{\hbar_p}{2} (\psi, \varphi' \left( \hbar_p - \hbar_p  (D_2)^{-2} R_2 + 2\hbar_p (D_2)^{-2} \nabla_l \nabla_{\bar l} \right) \nabla_m (\nabla_{\bar m} \chi') \phi)\\
&= \frac{\hbar_p^2}{2} (\psi, \varphi' \nabla_m (\nabla_{\bar m} \chi') \phi) + \hbar_p^2 (\psi, \varphi' (D_2)^{-2} \nabla_l \nabla_{\bar l} \nabla_m (\nabla_{\bar m} \chi') \phi) + O(\hbar_p^{5/2})\\
&= -\frac{\hbar_p^2}{2} (\psi, (\nabla_m \varphi') (\nabla_{\bar m} \chi') \phi) - \hbar_p  (\psi, \varphi' (D_2)^{-2} \nabla_m (\nabla_{\bar m} \chi') \phi)\\
& \quad + \hbar_p^2 (\psi, \varphi' (D_2)^{-2} \nabla_l \nabla_m (\nabla_{\bar l} \nabla_{\bar m} \chi') \phi) + O(\hbar_p^{5/2}).
}
Note that the second term of the last expression is exactly equal to $-\epsilon_2$.
This implies
\als{ \epsilon_2 &= -\frac{\hbar_p^2}{4} (\psi, (\nabla_m \varphi') (\nabla_{\bar m} \chi') \phi) + \epsilon'_2 + O(\hbar_p^{5/2}),\\
\epsilon'_2 &= \frac{\hbar_p^2}{2} (\psi, \varphi' (D_2)^{-2} \nabla_l \nabla_m (\nabla_{\bar l} \nabla_{\bar m} \chi') \phi)
}
Again using (\ref{asymptotic4}) to $\epsilon'_2$, we have
\als{ 
\epsilon'_2 &= \frac{\hbar_p^2}{4} (\psi, \varphi' \left( \hbar_p - \hbar_p  (D_2)^{-2} R_2 + 2\hbar_p (D_2)^{-2} \nabla_k \nabla_{\bar k} \right) \nabla_l \nabla_m (\nabla_{\bar l} \nabla_{\bar m} \chi') \phi)\\
&= \frac{\hbar_p^3}{2} (\psi, \varphi'  (D_2)^{-2} \nabla_k \nabla_{\bar k} \nabla_l \nabla_m (\nabla_{\bar l} \nabla_{\bar m} \chi') \phi) + O(\hbar_p^3)\\
&= \frac{\hbar_p^3}{2} (\psi, \varphi'  (D_2)^{-2} \nabla_k \nabla_l \nabla_{\bar k} \nabla_m (\nabla_{\bar l} \nabla_{\bar m} \chi') \phi) - \frac{\hbar_p^2}{2} (\psi, \varphi'  (D_2)^{-2} \nabla_l \nabla_m (\nabla_{\bar l} \nabla_{\bar m} \chi') \phi) + O(\hbar_p^3)\\
&= \frac{\hbar_p^3}{2} (\psi, \varphi'  (D_2)^{-2} \nabla_k \nabla_l \nabla_m (\nabla_{\bar k} \nabla_{\bar l} \nabla_{\bar m} \chi') \phi) - 2\epsilon'_2 + O(\hbar_p^3).
}
Similar to (\ref{norm order 1}), we can find $|\nabla_k \nabla_l \nabla_m \phi|= O(\hbar_p^{-3/2})$ and $|\nabla_l \nabla_m \phi|= O(\hbar_p^{-1})$.
This implies $\epsilon'_2 = O(\hbar_p^{5/2})$.
Therefore, we obtain
\eq{
    \epsilon = - \frac{\hbar_p^2}{4}(\psi, \varphi' R_2 \chi' \phi) -\frac{\hbar_p^2}{4} (\psi, (\nabla_m \varphi') (\nabla_{\bar m} \chi') \phi) + O(\hbar_p^{5/2}).
}
From (\ref{R_i}) and (\ref{gamma matrices identities}), one finds
\als{
    \epsilon &= \hbar_p^2 (\psi,  (\nabla_m \varphi) ( R_{\bar{m} l k\bar{k}} - R^{E_2}_{\bar{m}l}) (\nabla_{\bar l} \chi) \phi) + \frac{\hbar_p^2}{2} (\psi, (\nabla_m \nabla_l \varphi) (\nabla_{\bar m} \nabla_{\bar l} \chi) \phi) + O(\hbar_p^{5/2}).
}
This gives the coefficient $C_2(\varphi,\chi)$ of the asymptotic expansion (\ref{asymptotic exp}).

\subsection{Consistency check of the asymptotic expansion}
\label{Consistency check of the asymptotic expansion}
In this Appendix, we check that the asymptotic expansion 
(\ref{asym exp}) with (\ref{asymptotic exp}) derived 
in Appendix~\ref{asymptotic expansion} indeed satisfies the associativity of the Toeplitz operator product.
For $\varphi \in \Gamma({\rm Hom}(E_2,E_1))$,  
$\chi \in \Gamma({\rm Hom}(E_3,E_2))$ and
$\psi \in \Gamma({\rm Hom}(E_4,E_3))$, it should be true that
\eq{
\left( T_p^{(E_1,E_2)}(\varphi)T_p^{(E_2,E_3)}(\chi) \right)T_p^{(E_3,E_4)}(\psi) =  T_p^{(E_1,E_2)}(\varphi) \left(T_p^{(E_2,E_3)}(\chi) T_p^{(E_3,E_4)}(\psi)\right) .
\label{matrix associativity}
}
This imposes a condition
\als{
\sum_{i=0}^j
C_{j-i}\left(C_{i}(\varphi ,\chi),\psi \right)
-C_{i}\left(\varphi, C_{j-i}(\chi,\psi) \right) = 0,
\label{constraints for C}
}
for all $j\in \mathbb{Z}_{\ge 0}$.

We will check that the conditions (\ref{constraints for C}) for $j=0,1,2$ are satisfied by $C_0,C_1,C_2$ given in (\ref{asymptotic exp}). 
The condition for $j=0$ is satisfied from the associativity of the linear maps,
\als{
&C_{0}\left(C_0(\varphi,\chi),\psi \right)-
C_{0}\left(\varphi,C_{0}(\chi,\psi)\right)
=(\varphi \chi )\psi
-\varphi (\chi \psi) = 0.
}
For $j=1$, the left-hand side of (\ref{constraints for C}) is given by
\als{
&\sum_{i=0}^1
C_{1-i}\left(C_{i}(\varphi ,\chi),\psi \right)
-C_{i}\left(\varphi, C_{1-i}(\chi,\psi) \right)
\\ 
&= - \nabla_m (\varphi \chi) (\nabla_{\bar m} \psi)
+ \varphi (\nabla_m \chi)(\nabla_{\bar m} \psi)
- (\nabla_{m} \varphi)(\nabla_{\bar m} \chi) \psi
+ (\nabla_m \varphi)\nabla_{\bar m}(\chi\psi).
}
This is vanishing because of the Leibniz rule of the covariant derivatives.
Similarly, the condition for $j=2$ is also satisfied,
\als{
&\sum_{i=0}^2
C_{2-i}\left(C_{i}(\varphi ,\chi),\psi \right)
-C_{i}\left(\varphi, C_{2-i}(\chi,\psi) \right)
\\
&=
-(\nabla_m \varphi) \chi  R^{E_3}_{\bar{m}l} (\nabla_{\bar l} \psi) + (\nabla_m \varphi) R^{E_2}_{\bar{m}l} \chi (\nabla_{\bar l}\psi) - (\nabla_m \varphi) ([\nabla_{\bar m},\nabla_l] \chi) \nabla_{\bar l} \psi
\\
&=0.
}
Thus, the asymptotic expansion given in (\ref{asym exp}) and (\ref{asymptotic exp}) is consistent with the associativity condition (\ref{matrix associativity}) up to $\hbar_{p}^{2}$.

\subsection{Trace of Toeplitz operators}
\label{trace}
In this Appendix, we will show (\ref{trace property}).

First, by using the Schwartz kernel representation, the trace of $T_p^{(E_1,E_1)}(\varphi)$ is expressed as
\eq{
\label{trace schwarz kernel}
\Tr T_p^{(E_1,E_1)}(\varphi) = \int_M \mu(x) \mathrm{tr}_{S_c \otimes E_1} \left( B(x,x) \varphi(x)\right)
}
where $B(x,y)$ is the Bergman kernel defined by
\eq{
(\Pi_1 \psi) (x) = \int_M \mu(y) B(x,y) \psi(y) 
}
for any $\psi \in \Gamma(S_c \otimes L^{\otimes p} \otimes E_1)$.
In \cite{asym}, it is shown that the Bergmann Kernel has the following 
large-$p$ asymptotic expansion ,
\begin{align}
\label{B(x,x)}
B(x,x) = (2\pi \hbar_p)^{-n} P {\bf 1}_{E_1} + O(p^{n-1}),
\end{align}
where $P$ is the projection onto the zero mode component $\ket{+}^{\otimes n}$ of the fiber of $S$.
By plugging (\ref{B(x,x)}) into (\ref{trace schwarz kernel}), we obtain (\ref{trace property}).

\subsection{General properties of the Laplace operator on $\Gamma(E)$}
\label{Laplace operator}

In this Appendix, we show that the Bochner Laplacian defined in (\ref{def of Laplacian}) can be expressed as
\eq{ 
\Delta^{(E)}\varphi = - g^{\mu \nu} \nabla_{\mu} \nabla_{\nu}\varphi = - \{X^A, \{X^{A},\varphi\}\},}
where $\{ \cdot , \cdot\}$ is the generalized Poisson bracket (\ref{generalized Poisson bracket}) and $X^A$ is an isometric embedding function.

From the definition of the generalized Poisson bracket (\ref{generalized Poisson bracket}), we have
\als{- \{X^A,\{X^A,\varphi \}\}&= -W^{\alpha\beta}W^{\gamma\delta}
(\partial_{\alpha}X^A)\nabla_{\beta}[(\partial_{\gamma}X^A)(\nabla_{\delta}\varphi)]\\
&=
-W^{\alpha\beta}W^{\gamma\delta}(\partial_{\alpha}X^A)[(\nabla_{\beta}\partial_{\gamma}X^A)(\nabla_{\delta}\varphi) +(\partial_{\gamma}X^A)(\nabla_{\beta}\nabla_{\delta}\varphi) ]\\
&=
-W^{\alpha\beta}W^{\gamma\delta}[\nabla_{\beta} ( (\partial_{\alpha}X^A)
(\partial_{\gamma}X^A) ) (\nabla_{\delta}\varphi)-(\nabla_{\beta}\partial_{\alpha}X^A)(\partial_{\gamma}X^A)(\nabla_{\delta}\varphi)\\
&\qquad +(\partial_{\alpha}X^A)(\partial_{\gamma}X^A)(\nabla_{\beta}
\nabla_{\delta}\varphi) ]\\
&=-W^{\alpha\beta}W^{\gamma\delta}[(\nabla_{\beta}g_{\alpha\gamma})
(\nabla_{\delta}\varphi)-(\nabla_{\beta}\partial_{\alpha}X^A)
(\partial_{\gamma}X^A)(\nabla_{\delta}\varphi)
+g_{\alpha\gamma}(\nabla_{\beta}\nabla_{\delta}\varphi) ]
\\
&=-W^{\alpha\beta}W^{\gamma\delta}g_{\alpha\gamma}(\nabla_{\beta}
\nabla_{\delta}\varphi)
\\
&=-g^{\beta\delta}(\nabla_{\beta}\nabla_{\delta}\varphi).
\label{siki henkei}
}
Here, we used $\nabla W =0$, which follows from the general properties of the K\"ahler structure, $\nabla g = \nabla J = \nabla \omega=0$.
In the last equality, we used $W^{\alpha\beta}W^{\gamma\delta}g_{\alpha\gamma} = g^{\beta \delta}$, which we can check using the local orthonormal frame.
Therefore, (\ref{laplacian embedding}) holds for any K\"ahler manifold $M$.

\subsection{Simplification of the zero mode equation}
\label{simplification}

In this appendix, we argue that the Dirac equation is reduced to a simpler differential equation of holomorphic sections.

The twisted spin-$c$ Dirac operator $\Gamma(S_c \otimes L^{\otimes p} \otimes E)$ over $M$ is given by (\ref{twisted Dirac}).
From $\Omega_{ml} = \Omega_{\bar{m} \bar{l}} = 0$, we have
\eq{ \Omega_{ab} \gamma_{(2n)}^a \gamma_{(2n)}^b = \Omega_{m \bar{l}} \gamma_{(2n)}^m \gamma_{(2n)}^{\bar l} + \Omega_{\bar{l} m} \gamma_{(2n)}^{\bar l} \gamma_{(2n)}^m = 2 \Omega_{m \bar{l}} \gamma_{(2n)}^m \gamma_{(2n)}^{\bar l} -2 \sum_{m=1}^n \Omega_{m \bar{m}},}
where we used $\Omega_{m \bar{l}} = - \Omega_{\bar{l} m}$ and $\{\gamma_{(2n)}^m ,\gamma_{(2n)}^{\bar l}\} = 2\delta_{ml} I_{2^n}$ in the last equality.
As shown in Appendix \ref{Vanishing}, the zero mode $\psi$ is of the form $\psi = f \ket{+}^{\otimes n}$, where $f$ is a section of $L^{\otimes p} \otimes E$.
From (\ref{gamma matrices identities}), we then have
\als{&D\psi = \im  \bar{w}_m{}^{\bar{\mu}} \gamma_{(2n)}^{\bar{m}}\ket{+}^{\otimes n} \left(\partial_{\bar{\mu}} + p A^L_{\bar \mu} + A^E_{\bar \mu} \right) f = 0 \\
& \Rightarrow \quad {}^{\forall} m \in \{1,\cdots, n\}: \quad  \bar{w}_m {}^{\bar{\mu}} \left(\partial_{\bar{\mu}} + p A^L_{\bar \mu} + A^E_{\bar \mu} \right) f = 0\\
& \Rightarrow \quad \left(\partial_{\bar{\mu}} + p A^L_{\bar \mu} + A^E_{\bar \mu} \right) f = 0.}
This indicates that $f$ is a holomorphic section of $L^{\otimes p} \otimes E$.

\section{Proofs and formulas for $CP^n$}

\subsection{Integration formula for $CP^n$}
\label{integral formula}

In this Appendix, we calculate
\eq{I_{s,t,p} := \int_{CP^n} \mu \,  \frac{\prod_{i=1}^n (\bar{z}^i)^{s_i} (z^i)^{t_i}}{(1+|z|^2)^{p}},}
which is a typical integral appearing in our discussion of $CP^n$.
Here, $s=(s_1,s_2,\cdots,s_n), \, t=(t_1,t_2,\cdots,t_n) \in (\mathbb{Z}_{\ge 0})^n$ and $p\in \mathbb{Z}$.
The result is
\eq{\label{I_s,p}
I_{s,t,p} = I_{s,p} \delta_{s,t}, \quad 
I_{s,p}= \frac{(2\pi)^n(p - \sum_{i=1}^n s_i)! \prod_{i=1}^n (s_i !)}{(p+n)!}.}
Here, the Kronecker delta is defined as $\delta_{s,t}:= \prod_{i=1}^n \delta_{s_i,t_i}$ and the factor $I_{s,p}$ is convergent if and only if 
\eq{
\label{convergent}
\sum_{i=1}^n s_i < p +1.}

Now, let us begin the proof.
First, since $CP^n \setminus U_{\alpha}$ has zero measure, the integral over $CP^n$ is computed in a single patch:
\eq{I_{s,t,p} = \int_{\mathbb{R}^{2n}} \frac{\prod_{i=1}^n \left(\frac{x^{2i-1}- \im x^{2i}}{\sqrt{2}}\right)^{s_i} \left(\frac{x^{2i-1}+ \im x^{2i}}{\sqrt{2}}\right)^{t_i}}{(1+\frac{|x|^2}{2})^{p+n+1}} \dif x^1 \dif x^2 \cdots \dif x^{2n}.}
Here, we are using a real coordinates $x=(x^1, x^2,\cdots,x^{2n})$ defined by
\eq{x^{2\mu-1} = \frac{z^{\mu} + \bar{z}^{\mu}}{\sqrt{2}}, \quad x^{2\mu} = \frac{z^{\mu} - \bar{z}^{\mu}}{\sqrt{2}\im.}}
We can employ the angular coordinates $(\rho_i,\theta_i)\in [0,\infty) \times [0,2\pi)$ such that \eq{x^{2i-1} = \sqrt{2} \, \rho_i \cos \theta_i, \ x^{2i} = \sqrt{2} \, \rho_i \sin \theta_i.}
This gives
\eq{I_{s,t,p} = \prod_{i=1}^n\left(2\int_{0}^{\infty} \rho_i \dif \rho_i \int_0^{2\pi} \dif \theta_i\right) \, \frac{\prod_{i=1}^n \left(\rho_i \e^{\im \theta_i}\right)^{s_i} \left( \rho_i \e^{- \im \theta_i} \right)^{t_i}}{(1+\sum_{i=1}^n \rho_i^2)^{p+n+1}}.}
The angular integrals give a factor $\delta_{s,t}$.
Then, we obtain $I_{s,t,p} = I_{s,p} \delta_{s,t}$ where
\eq{I_{s,p} = (4\pi)^n \int_{[0,\infty)^n} \frac{\dif \rho_1 \dif \rho_2 \cdots \dif \rho_n}{(1+\sum_{i=1}^n \rho_i^2)^{p+n+1}} \prod_{i=1}^n \rho_i^{2s_i+1}.}
We can use the spherical coordinates $(\rho,\phi_1 ,\phi_2,\cdots,\phi_{n-1})\in [0,\infty) \times [0,\pi/2]^{n-1}$ given by
\eq{\rho_1 = \rho \cos \phi_1, \quad \rho_2 = \rho \sin \phi_1 \cos \phi_2, \quad \cdots,  \quad \rho_{n-1} =
\rho \left( \prod_{i=1}^{n-2}\sin \phi_i\right) \cos \phi_{n-1},  \quad \rho_n =
\rho \prod_{i=1}^{n-1}\sin \phi_i,}
and we obtain
\eq{I_{s,p} 
= (4\pi)^n \int_0^{\infty} \dif \rho \frac{ \rho^{2\sum_{i=1}^n (s_i +1) -1}}{(1+\rho^2)^{p+n+1}} \prod_{i=1}^{n-1} \left(\int_0^{\pi/2} \dif \phi_i \sin^{2\sum_{j=i+1}^n (s_j + 1) -1} (\phi_i) \cos^{2s_i +1} (\phi_i)\right).
}
Note that Beta function
\eq{ B(x,y) = 2\int_0^{\pi/2} \dif \phi \sin^{2x-1} \phi \, \cos^{2y-1} \phi  = 2 \int_0^{\infty} \dif \rho \frac{\rho^{2x-1}}{(1+\rho^2)^{x+y}}}
only converges for $\mathrm{Re}\,x,\mathrm{Re} \, y>0$.
Then, we can see that $I_{r,w}$ is convergent if and only if (\ref{convergent}) is satisfied and the value of $I_{s,p}$ is
\eq{
I_{s,p}= (2\pi)^n B \left(\sum_{i=1}^n (s_i +1), p +1 - \sum_{i=1}^n s_i \right) \prod_{i=1}^{n-1} B \left(\sum_{j=i+1}^n (s_j + 1), s_i+1 \right).
}
Using $B(x,y) = \Gamma(x)\Gamma(y)/ \Gamma(x+y)$ and $\Gamma(x+1) =x!$, we finally obtain (\ref{I_s,p}).

\subsection{Proof of (\ref{algebra of generators on sections})}
\label{differential operator commutation relation}
Here, we will show that the operator (\ref{generators on sections}) satisfies the commutation relation (\ref{algebra of generators on sections}).

We first show
\eq{
\label{Poisson bracket of embedding functions}
\{X^A,X^B\}=- \sqrt{2} f_{ABC} X^C,
}
which is needed in the proof of (\ref{algebra of generators on sections}).
In the complex coordinates, the Poisson tensor is given by $W^{\mu\bar{\nu}} = - W^{\bar{\nu}\mu} = - \im g^{\mu\bar{\nu}}$ and it gives
\eq{
\label{Poisson bracket CP^n}
\{X^A,X^B\} = - \im g^{\mu\bar{\nu}} [(\partial_{\mu} X^A)(\partial_{\bar{\nu}} X^B) - (A \leftrightarrow B) ].}
From (\ref{X^A}) and (\ref{zeta}), the embedding function can be written as
\eq{ X^A = -\frac{\sqrt{2}}{1+|z|^2} \left(\bar{z}^{\mu} (T_A)_{\mu\nu} z^{\nu} + (T_A)_{\mu\, n+1} \bar{z}^{\mu} + (T_A)_{n+1 \, \mu} z^{\mu} + (T_A)_{n+1 \, n+1}\right).}
By using this expression, we have
\eq{ \partial_{\mu}X^A =  - \frac{\bar{z}^{\mu} X^A}{1+|z|^2} - \frac{\sqrt{2}(\zeta^{\dagger} T_A)_{\mu}}{\sqrt{1+|z|^2}}, \quad \partial_{\bar{\nu}}X^A =  - \frac{z^{\nu} X^A}{1+|z|^2} - \frac{\sqrt{2}(T_A \zeta)_{\nu}}{\sqrt{1+|z|^2}}.
}
Also using (\ref{metric inverse}), we obtain
\als{
\label{dX^A}
&g^{\mu\bar{\nu}} \partial_{\mu}X^A = \sqrt{2(1+|z|^2)}\left( (\zeta^{\dagger} T_A)_{n+1} \bar{z}^{\nu} - (\zeta^{\dagger} T_A)_{\nu} \right), \\ &g^{\bar{\nu} \mu} \partial_{\bar{\nu}}X^A =  \sqrt{2(1+|z|^2)} \left( (T_A \zeta)_{n+1} z^{\mu} - (T_A \zeta)_{\mu} \right).
}
Thus, we have
\als{ \label{Poisson bracket X^A}
\{X^A,X^B\} &= \im \sqrt{2(1+|z|^2)}\left( (\zeta^{\dagger} T_A)_{n+1} \bar{z}^{\nu} - (\zeta^{\dagger} T_A)_{\nu} \right) \left(  \frac{z^{\nu} X^B}{1+|z|^2} + \frac{\sqrt{2}(T_B \zeta)_{\nu}}{\sqrt{1+|z|^2}} \right) - (A \leftrightarrow B)\\
&= \im 2\zeta^{\dagger} [T_A,T_B] \zeta.
}
Using (\ref{T_A algebra}) and (\ref{X^A}), we obtain (\ref{Poisson bracket of embedding functions}).

Let us prove (\ref{algebra of generators on sections}).
From the definition (\ref{generators on sections}), we have
\eq{ [\mathcal{L}^{(q)}_A,\mathcal{L}^{(q)}_B] f^{(q)} = - \frac{1}{2}\{X^A,\{X^B, f^{(q)}\} \} + \frac{1}{2}\{X^B,\{X^A, f^{(q)}\} \} - \im  q\{X^A,X^B\}f^{(q)} .}
Using the definition of the generalized Poisson bracket, we calculate as follows,
\als{ 2[\mathcal{L}^{(q)}_A,\mathcal{L}^{(q)}_B] f^{(q)} &= - W^{\alpha\beta}W^{\gamma \delta} (\partial_{\alpha}X^A) \nabla_{\beta} [(\partial_{\gamma} X^B) (\nabla_{\delta} f^{(q)})]\\
&\quad + W^{\alpha\beta}W^{\gamma \delta} (\partial_{\alpha}X^B) \nabla_{\beta} [(\partial_{\gamma} X^A) (\nabla_{\delta} f^{(q)})] - \im  2q\{X^A,X^B\}f^{(q)}\\
&= - \im  q\{X^A,X^B\}f^{(q)} - (W^{\alpha\beta}W^{\gamma \delta} - W^{\gamma\beta}W^{\alpha \delta}) (\nabla_{\beta}[(\partial_{\alpha}X^A)  (\partial_{\gamma} X^B)]) (\nabla_{\delta} f^{(q)}).
}
Here, we used $[\nabla_{\beta},\nabla_{\delta}] f^{(q)} = - \im q\omega_{\beta\delta} f^{(q)}$ and $\omega_{\mu\nu}W^{\mu\rho} = \delta_{\nu}^{\rho}$.
By using $W^{\alpha\beta}W^{\gamma \delta} - W^{\gamma\beta}W^{\alpha \delta} = W^{\alpha\gamma} W^{\beta\delta}$, which we can check in the orthonormal coordinates, we obtain
\als{ [\mathcal{L}^{(q)}_A,\mathcal{L}^{(q)}_B] f^{(q)} &=  - \frac{1}{2} W^{\alpha\gamma} W^{\beta\delta} (\nabla_{\beta}[(\partial_{\alpha}X^A)  (\partial_{\gamma} X^B)]) (\nabla_{\delta} f^{(q)}) - \im  \frac{q}{2}\{X^A,X^B\}f^{(q)}\\
&= - \frac{1}{2}\{\{X^A,X^B\}, f^{(q)}\} - \im  \frac{q}{2}\{X^A,X^B\}f^{(q)}.
}
Therefore, using (\ref{Poisson bracket of embedding functions}) and (\ref{generators on sections}), we have shown the relation (\ref{algebra of generators on sections}).

\subsection{Direct calculation of (\ref{E_k CP^n})}
\label{cp^n laplacian eigenvalues}

Let us evaluate $\Delta^{(q)} f_{k,I}^{(q)}$.
First, by the definition of $\Delta^{(q)}$, we have
\eq{ \Delta^{(q)} = -g^{\mu\bar{\nu}} \left( D^{(q)}_{\mu} D^{(q)}_{\bar \nu} + D^{(q)}_{\bar \nu}D^{(q)}_{\mu} \right), }
The covariant derivatives on $f_{\bm{\sigma}_{k+q}, \bm{\tau}_{k}}^{(q)}$ defined in (\ref{basis of monopole section}) are given by
\als{
\label{Df}
D^{(q)}_{\mu} f_{\bm{\sigma}_{k+q}, \bm{\tau}_{k}}^{(q)} =& (\partial_{\mu} + q A^L_{\mu})f_{\bm{\sigma}_{k+q}, \bm{\tau}_{k}}^{(q)} = \left( \sum_{a=1}^{k+q}\frac{\delta_{\mu,\sigma_a}}{z^{\sigma_a}} - (k+q) \frac{\bar{z}^{\mu}}{1+|z|^2}\right) f_{\bm{\sigma}_{k+q}, \bm{\tau}_{k}}^{(q)},\\
D^{(q)}_{\bar \nu} f_{\bm{\sigma}_{k+q}, \bm{\tau}_{k}}^{(q)} =& (\partial_{\bar \nu} + q A^L_{\bar \nu}) f_{\bm{\sigma}_{k+q}, \bm{\tau}_{k}}^{(q)} =\left( \sum_{b=1}^{k}\frac{\delta_{\nu,\tau_b}}{\bar{z}^{\tau_b}} - k \frac{z^{\nu}}{1+|z|^2}\right)f_{\bm{\sigma}_{k+q}, \bm{\tau}_{k}}^{(q)}.
}
Here, we set $z^{n+1} = \bar{z}^{n+1}=1$.
Thus, we have
\als{ \Delta^{(q)} f_{\bm{\sigma}_{k+q}, \bm{\tau}_{k}}^{(q)} &= -2 g^{\mu\bar{\nu}} \left[\left( \sum_{a=1}^{k+q}\frac{\delta_{\mu,\sigma_a}}{z^{\sigma_a}} - (k+q) \frac{\bar{z}^{\mu}}{1+|z|^2}\right) \left( \sum_{b=1}^{k}\frac{\delta_{\nu,\tau_b}}{\bar{z}^{\tau_b}} - k \frac{z^{\nu}}{1+|z|^2}\right) - \left(k+\frac{q}{2} \right) g_{\mu \bar{\nu}} \right] f_{\bm{\sigma}_{k+q}, \bm{\tau}_{k}}^{(q)}. }
By using (\ref{metric inverse}) and (\ref{f_k,I}), we obtain
\als{
\label{Laplacian eigenvalues}
\Delta^{(q)} f_{k,I}^{(q)} = 2 \left( k(k+q) + n\left(k+\frac{q}{2} \right) \right) f_{k,I}^{(q)}.
}
Here, we used the traceless property $\sum_{\sigma_a,\tau_b} c^{(q)}_{I, \bm{\sigma}_{k+q}, \bm{\tau}_k} \delta_{\sigma_a,\tau_b} =0$.
By comparing (\ref{Laplacian eigenvalues}) with (\ref{Laplacian and Casimir}), we find (\ref{E_k CP^n}).

\subsection{Proof of (\ref{L_A correspondence})}
\label{su(n) generator correspondence}

In this appendix, we give a proof of the important identity (\ref{L_A correspondence}).

Using (\ref{dX^A}) and (\ref{Df}), we have
\als{ \label{L_A f}
\mathcal{L}_A^{(q)} f_{\bm{\sigma}_{k+q}, \bm{\tau}_{k}}^{(q)}
&= \left( - \sum_{a=1}^{p+q} \frac{(T_A \zeta)_{\sigma_a}}{\zeta^{\sigma_a}}
+ \sum_{b=1}^{p} \frac{(\zeta^{\dagger}T_A )_{\tau_b}}{\bar{\zeta}^{\tau_b}}\right)
f_{\bm{\sigma}_{k+q}, \bm{\tau}_{k}}^{(q)}\\
& =  - \sum_{i,j=1}^{n+1}(T_A)_{i j} n_i(\bm{\sigma}_{k+q}) f_{\bm{\sigma}_{k+q} \ominus i \oplus j, \bm{\tau}_{k}}^{(q)}
+ \sum_{i,j=1}^{n+1}(T_A)_{i j} n_j (\bm{\tau}_{k}) f_{\bm{\sigma}_{k+q} , \bm{\tau}_{k} \ominus j \oplus i}^{(q)}.
}
Here, $n_{i}(\bm{\alpha}_p)$ is the number of components of $\bm{\alpha}_p$ equal to $i$ and $\ominus$ is the inverse operation of $\oplus$, namely, $\bm{\tau}_{k} \ominus j = (\tau_1,\cdots, \tau_{b-1}, \tau_{b+1}, \cdots, \tau_k)$ for $j= \tau_b$.
We calculate the Toeplitz operator of the above object as
\als{
\label{T(Lf)}
&T_p(\mathcal{L}_A^{(q)} f_{\bm{\sigma}_{k+q}, \bm{\tau}_{k}}^{(q)})_{\bm{\alpha}_{p+q}, \bm{\beta}_p}\\
& =  -  \sum_{i,j=1}^{n+1}(T_A)_{i j} n_i(\bm{\sigma}_{k+q}) T_p(f_{\bm{\sigma}_{k+q} \ominus i \oplus j, \bm{\tau}_{k}}^{(q)})_{\bm{\alpha}_{p+q}, \bm{\beta}_p} +  \sum_{i,j=1}^{n+1}(T_A)_{ij} n_j (\bm{\tau}_{k}) T_p(f_{\bm{\sigma}_{k+q} , \bm{\tau}_{k} \ominus j \oplus i}^{(q)})_{\bm{\alpha}_{p+q}, \bm{\beta}_p}\\
& =  \frac{c^{(p+q)}_{\bm{\alpha}_{p+q}} c^{(p)}_{\bm{\beta}_p}}{\left(c^{(p+q+k)}_{\bm{\alpha}_{p+q} \oplus \bm{\tau}_k} \right)^2} \sum_{i,j=1}^{n+1}(T_A)_{i j} \delta_{\bm{\alpha}_{p+q} \oplus \bm{\tau}_k \oplus i, \bm{\beta}_{p} \oplus \bm{\sigma}_{k+q} \oplus j} \left[- n_i(\bm{\sigma}_{k+q}) + 
\frac{ n_j (\bm{\tau}_{k})  (n_i(\bm{\alpha}_{p+q} \oplus \bm{\tau}_k) +1 )}{n_j(\bm{\alpha}_{p+q} \oplus \bm{\tau}_k) + \delta_{i,j}} \right],
}
where we used (\ref{T(f)}) and (\ref{normalization constant}).
On the other hand, $L_A \circ T_p (f_{\bm{\sigma}_{k+q}, \bm{\tau}_{k}}^{(q)})$ is given by
\als{ 
\label{LT(f)}
(L_A \circ T_p (f_{\bm{\sigma}_{k+q}, \bm{\tau}_{k}}^{(q)}))_{\bm{\alpha}_{p+q}, \bm{\beta}_p}
&= \frac{c^{(p+q)}_{\bm{\alpha}_{p+q}} c^{(p)}_{\bm{\beta}_p}}{\left(c^{(p+q+k)}_{\bm{\alpha}_{p+q} \oplus \bm{\tau}_k} \right)^2} \sum_{i,j=1}^{n+1} (T_A)_{ij} \delta_{\bm{\alpha}_{p+q} \oplus \bm{\tau}_k \oplus i, \bm{\beta}_{p} \oplus \bm{\sigma}_{k+q} \oplus j} \\
& \quad \times \left( -  \frac{(n_j(\bm{\alpha}_{p+q}) + \delta_{i,j})(n_i(\bm{\alpha}_{p+q} \oplus \bm{\tau}_k)+1)}{n_j(\bm{\alpha}_{p+q} \oplus \bm{\tau}_k) + \delta_{i,j}} + n_{i}(\bm{\beta}_p) + \delta_{i,j} \right).
}
Here, we used (\ref{T(f)}), (\ref{normalization constant}) and the following expression of $L_A^{(p)}$,
\eq{
(L^{(p)}_A)_{\bm{\alpha}_p, \bm{\beta}_p} = - \frac{c^{(p)}_{\bm{\beta}_p}}{c^{(p)}_{\bm{\alpha}_p}} \sum_{i,j=1}^{n+1}(T_A)_{ij} (n_{i}(\bm{\alpha}_p) +1)  \delta_{\bm{\alpha}_p \oplus i,\bm{\beta}_p \oplus j},
}
which follows from (\ref{T(X^A) CP^n}) and (\ref{embedding regularization}). 
Comparing (\ref{T(Lf)}) with (\ref{LT(f)}), we find $T_p (\mathcal{L}_A^{(q)} f_{\bm{\sigma}_{k+q}, \bm{\tau}_{k}}^{(q)}) = L_A \circ T_p (f_{\bm{\sigma}_{k+q}, \bm{\tau}_{k}}^{(q)})$, which implies (\ref{L_A correspondence}).

\subsection{Proof of (\ref{cp^n inner product correspondence})}
\label{cp^n inner product eigenfunctions}
In this appendix, we prove (\ref{cp^n inner product correspondence}).

Let us start with (\ref{orthogonality of matrix})
for $k,k' \le p$.
For fixed $k \le p$, we first show that the proportional factor does not depend on $I$ labelling the different weights of the eigenstates.
Let $\{H_a\}_{a=1}^n$ be a basis of Cartan subalgebra of $\mathfrak{su}(n+1)$, that is, a set of mutually commuting elements in $\{T_A\}_{A=1}^{n^2 + 2n}$.
Then, there exists a complete basis of $\mathfrak{su}(n+1)$ called Cartan-Weyl basis $\{H_a, E_{\alpha}\}$ satisfying
\eq{
\label{Cartan-Weyl basis}
[H_a,H_b] = 0, \quad [H_a,E_{\pm \alpha}] = \pm \alpha_a E_{\pm \alpha}, \quad [E_{\alpha},E_{-\alpha}] = \sum_a \alpha_a H_a, \quad E_{\alpha}^{\dagger} = E_{-\alpha}.
}
Here, $\alpha$ runs over all roots of $\mathfrak{su}(n+1)$.
Now, let us consider its irreducible representations $\rho_1 : \mathfrak{su}(n+1) \to \mathrm{End}(V_1)$ and $\rho_2 : \mathfrak{su}(n+1) \to \mathrm{End}(V_2)$, where
\al{
&V_1 = \mathrm{Span}_{\mathbb{C}} (\{f_{k,I}\}), \hspace{11mm} \rho_1(T_A) = \mathcal{L}^{(q)}_A, \\
&V_2 = \mathrm{Span}_{\mathbb{C}} (\{T_p(f_{k,I})\}), \quad \rho_2(T_A) = L_A \circ.
}
for fixed 
Here, $k$ is fixed and $V_1$ and $V_2$ shall be generated by running the subscript $I$ over all weights.
We take the label $I$ as the $n$-dimensional vector $I = (I_1,I_2,\cdots,I_n)$ such that
\eq{\rho_1(H_a) f_{k,I} = I_a f_{k,I}.}
In this notation, the correspondence (\ref{L_A correspondence}) implies
\eq{
\label{representation correspondence}
\rho_2(v) T_p(f_{k,I}) = T_p(\rho_1(v) f_{k,I}),
}
for any $v \in \mathfrak{su}(n+1)$.
Note that from (\ref{Cartan-Weyl basis}), we have
\eq{
\label{raising operator}
\rho_1(E_{\alpha}) f_{k,I} = N_{\alpha,I} f_{k,I+\alpha}
}
for a complex constant number $N_{\alpha,I}$.
We again act $\rho_1(E_{-\alpha})$ on both sides of (\ref{raising operator}) and obtain
\eq{
\label{C_alpha,I}
\rho_1(E_{-\alpha} E_{\alpha}) f_{k,I} = C_{\alpha,I} f_{k,I},
}
where $C_{\alpha,I}$ is given by
\eq{ C_{\alpha,I}  = (f_{k,I},\rho_1(E_{-\alpha}E_{\alpha}) f_{k,I}) = (\rho_1(E_{\alpha}) f_{k,I},\rho_1(E_{\alpha}) f_{k,I}) = |N_{\alpha,I}|^2.}
Here, we assumed that $f_{k,I}$ and $f_{k,I+\alpha}$ are both normalized.
From (\ref{representation correspondence}), (\ref{raising operator}) and (\ref{C_alpha,I}), we have
\als{
(T_p (f_{k,I+\alpha}), T_p (f_{k,I+\alpha})) &= |N_{\alpha,I}|^{-2} (\rho_2(E_{\alpha}) T_p (f_{k,I+\alpha}), \rho_2(E_{\alpha}) T_p (f_{k,I+\alpha}))\\
&=|N_{\alpha,I}|^{-2} (T_p (f_{k,I+\alpha}), \rho_2(E_{-\alpha} E_{\alpha}) T_p (f_{k,I+\alpha}))\\
&=|N_{\alpha,I}|^{-2} (T_p (f_{k,I+\alpha}), T_p (\rho_1(E_{-\alpha} E_{\alpha}) f_{k,I+\alpha}))\\
&= (T_p (f_{k,I}), T_p (f_{k,I})).
}
Since this holds for any $I$ and $\alpha$, we find
\eq{
(T_p (f_{k,I}), T_p (f_{k,I})) = (T_p (f_{k,I'}), T_p (f_{k,I'})),
}
for general weights $I,I'$.

From the above argument, we only have to compute $(T_p (f_{k,I}), T_p (f_{k,I}))$ for a specific $I$.
Let us consider a particular element
\eq{f^{(q)}_{k, I}:= c^{(2k+q)}_{\bm{1}_{k+q} \oplus \bm{2}_{k}} (\zeta^{1})^{k+q} (\bar{\zeta}^{2})^k.}
Here, we introduced $\bm{1}_{k+q} = (1,1,\cdots,1)$ and $\bm{2}_{k} = (2,2,\cdots,2)$.
From (\ref{normalization constant}), the normalization constant is given by $c^{(2k+q)}_{\bm{1}_{k+q} \oplus \bm{2}_{k}} := \sqrt{\frac{(2k+q+n)!}{(2\pi)^{n}k!(k+q)!}}$.
By using (\ref{T(f)}), we have
\eq{ T_p(f^{(q)}_{k, I})_{\bm{\alpha}_{p+q}, \bm{\beta}_p} = c^{(2k+q)}_{\bm{1}_{k+q} \oplus \bm{2}_{k}} \frac{c_{\bm{\alpha}_{p+q}}^{(p+q)} c_{\bm{\beta}_{p}}^{(p)}}{ \left(c_{\bm{\alpha}_{p+q} \oplus \bm{2}_{k}}^{(p+k+q)} \right)^2} \delta_{\bm{\alpha}_{p+q} \oplus \bm{2}_{k}, \bm{\beta}_p \oplus \bm{1}_{k+q}}
}
and the only non-vanishing components are
\eq{ T_p(f^{(q)}_{k, I})_{\bm{1}_{k+q} \oplus \bm{\rho}_{p-k}, \bm{2}_{k} \oplus \bm{\rho}_{p-k}} =  c^{(2k+q)}_{\bm{1}_{k+q} \oplus \bm{2}_{k}} \frac{c_{\bm{1}_{k+q} \oplus \bm{\rho}_{p-k}}^{(p+q)} c_{\bm{2}_{k} \oplus \bm{\rho}_{p-k}}^{(p)}}{ \left(c_{\bm{1}_{k+q} \oplus \bm{2}_{k} \oplus \bm{\rho}_{p-k}}^{(p+k+q)} \right)^2}.
}
For $p-k<0$, we see that such matrices should vanish.
Using (\ref{normalization constant}), we find
\eq{
\label{norm formula 1}
(T_p(f^{(q)}_{k, I}), T_p(f^{(q)}_{k, I}))
=\frac{(2k+q+n)!(p+q+n)!(p+n)!}{(2\pi)^n k! (k+q)!((p+q+k+n)!)^2} \sum_{\bm{\rho}_{p-k}} \frac{(n_1(\bm{\rho}_{p-k}) +k+q)!}{n_1(\bm{\rho}_{p-k})!} \frac{(n_2(\bm{\rho}_{p-k}) +k)!}{n_2(\bm{\rho}_{p-k})!}.
}
Let us set $a:= n_1(\bm{\rho}_{p-k})$ and $b:= n_2(\bm{\rho}_{p-k})$ which satisfy $0 \le a+b\le p-k$.
Here, for fixed $a$ and $b$, the number of possible configurations of $\bm{\rho}_{p-k}$ is $\frac{(p-k-a-b+n-2)!}{(n-2)!(p-k-a-b)!}$ for $n > 1$.
Thus, we have
\eq{
\label{n>1}
\sum_{\bm{\rho}_{p-k}} \frac{(n_1(\bm{\rho}_{p-k}) +k+q)!}{n_1(\bm{\rho}_{p-k})!} \frac{(n_2(\bm{\rho}_{p-k}) +k)!}{n_2(\bm{\rho}_{p-k})!} = \sum_{a=0}^{p-k} \sum_{b=0}^{p-k-a} \frac{(a+k+q)!}{a!} \frac{(b+k)!}{b!} \frac{(p-k-a-b+n-2)!}{(n-2)!(p-k-a-b)!},
}
for $n>1$.
Let us use the Chu-Vandermonde identity,
\eq{
\label{S_m formula}
\sum_{a=0}^m \frac{(a+i)!(j+m-a)!}{a!(m-a)!} = \frac{i!j!(i+j+m+1)!}{(i+j+1)! m!},
}
for any non-negative integers $m,i$ and $j$.
By applying this identity to (\ref{n>1}), we find
\eq{ \label{sum rho}
\sum_{\bm{\rho}_{p-k}} \frac{(n_1(\bm{\rho}_{p-k}) +k+q)!}{n_1(\bm{\rho}_{p-k})!} \frac{(n_2(\bm{\rho}_{p-k}) +k)!}{n_2(\bm{\rho}_{p-k})!} = \frac{k!(k+q)!(p+q+k+n)!}{(2k+q+n)!(p-k)!}.
}
For $n=1$, we have
\als{\sum_{\bm{\rho}_{p-k}} \frac{(n_1(\bm{\rho}_{p-k}) +k+q)!}{n_1(\bm{\rho}_{p-k})!} \frac{(n_2(\bm{\rho}_{p-k}) +k)!}{n_2(\bm{\rho}_{p-k})!} &= \sum_{a=0}^{p-k} \frac{(a+k+q)!}{a!} \frac{(p-a)!}{(p-k-a)!}\\
&= \frac{k!(k+q)!(p+q+k+1)!}{(2k+q+1)!(p-k)!},
}
and thus (\ref{sum rho}) holds for any $n \in \mathbb{N}$.
By plugging (\ref{sum rho}) into (\ref{norm formula 1}), we obtain (\ref{cp^n inner product correspondence}).

\section{Proofs and formulas for $T^{2n}$}

\subsection{Integration formula for $T^{2n}$}
\label{integration formula 2}
In this Appendix, we explicitly calculate 
\eq{
\label{def of I_m,i,j}
I_{m,i_m,j_m}^{(a,b)} := \int_{0}^{2\pi l_{2m-1}} \dif x^{2m-1}\int_{0}^{2\pi l_{2m}} \dif x^{2m} (f^{(p)}_{i_m})^* \e^{\im \frac{a x^{2m-1}}{l_{2m-1}}} \e^{\im \frac{b x^{2m}}{l_{2m}}} f^{(p)}_{j_m}.}
Here, $a,b \in \mathbb{Z}$ and $f^{(p)}_{i_m}$ is defined in (\ref{torus zero mode 3}).

By plugging (\ref{torus zero mode 3}) into (\ref{def of I_m,i,j}), we have
\als{
I_{m,i_m,j_m}^{(a,b)} &= \left( \frac{kp}{4 \pi^3 l_{2m}^2}\right)^{1/2} \sum_{l,l' \in \mathbb{Z}} 
\e^{\im\pi \left(l + \frac{i_m}{pq_m} \right)^2 pq_m\tau_m}
\e^{\im\pi \left(l' + \frac{j_m}{pq_m} \right)^2 pq_m\tau_m}\\
& \quad \times\int_{0}^{2\pi l_{2m-1}} \dif x^{2m-1} \e^{-kp (x^{2m-1})^2} \e^{\left(l + l' + \frac{i_m + j_m}{pq_m} \right) \frac{pq_m x^{2m-1}}{l_{2m}}} \e^{\im \frac{a x^{2m-1}}{l_{2m-1}}}\\
& \quad \times \int_{0}^{2\pi l_{2m}} \dif x^{2m} \e^{- \im \left(l -l' + \frac{i_m - j_m - b}{pq_m} \right) \frac{pq_m x^{2m}}{l_{2m}}}.
}
Then, performing the integral of $x^{2m}$ and taking the summation of $l'$, we obtain
\als{
I_{m,i_m,j_m}^{(a,b)} &= \left( \frac{kp}{\pi}\right)^{1/2} \delta^{(\mathrm{mod} \ pq_m)}_{i_m - j_m - b,0} \ \e^{- \frac{1}{4kp} \left( \frac{a^2}{l_{2m-1}^2} + \im \frac{2ab}{l_{2m-1} l_{2m}}+  \frac{b^2}{l_{2m}^2} \right)} \e^{\im\frac{2\pi a i_m}{pq_m}} \\
& \quad \times \sum_{l \in \mathbb{Z}} \int_{0}^{2\pi l_{2m-1}} \dif x^{2m-1} \e^{-kp \left(x^{2m-1} - 2\pi l_{2m-1} \left( l + \frac{i_m}{pq_m}\right) - \frac{\im}{2kp}\left( \frac{a}{l_{2m-1}} + \im \frac{b}{l_{2m}}\right) \right)^2}.
}
Here, we defined
\eq{\delta^{(\mathrm{mod} \ n)}_{a,b} = \begin{cases}
    1 & (a-b \in n\mathbb{Z})\\
    0 & (\text{otherwise})
\end{cases}.
}
By shifting the coordinate $x^{2m-1} \mapsto x^{2m-1}+2\pi l_{2m-1}l$, we can convert the summation of $l$ into extending the integration range to $\mathbb{R}$.
This yields the usual Gaussian integral and we obtain
\als{
\label{I_m components}
I_{m,i_m,j_m}^{(a,b)} &= \e^{- \frac{1}{4kp} \left( \frac{a^2}{l_{2m-1}^2} + \im \frac{2ab}{l_{2m-1} l_{2m}}+  \frac{b^2}{l_{2m}^2} \right)} \e^{\im\frac{2\pi a i_m}{pq_m}} \delta^{(\mathrm{mod} \ pq_m)}_{i_m - j_m - b,0}. 
}
For $a=b=0$, we can see that
\als{
I_{m,i_m,j_m}^{(0,0)} &= \delta^{(\mathrm{mod} \ pq_m)}_{i_m - j_m,0},
}
which means the orthonormality (\ref{orthonormality of f_i_m}).
For $(a,b)=(1,0)$ and $(0,1)$, we can see that (\ref{I_m components}) can be written in terms of the clock and shift matrices (\ref{U,V}) as
\eq{ I_{m,i_m,j_m}^{(1,0)} = (U_{pq_m})_{i_m,j_m}, \quad  I_{m,i_m,j_m}^{(0,1)} = (V_{pq_m})_{i_m,j_m}.}
Similarly, for general $a$ and $b$, we have
\eq{ I_{m,i_m,j_m}^{(a,b)} = \e^{- \frac{1}{4kp} \left( \frac{a^2-a}{l_{2m-1}^2} + \im \frac{2ab}{l_{2m-1} l_{2m}}+  \frac{b^2-b}{l_{2m}^2} \right)} ((U_{pq_m})^a (V_{pq_m})^b)_{i_m,j_m}.}

\subsection{Proof of (\ref{fuzzy torus T_p(f)})}
\label{eigenstate products}

In this appendix, we give a derivation of (\ref{fuzzy torus T_p(f)}).

To show (\ref{fuzzy torus T_p(f)}), we introduce the Jacobi theta function
\eq{ \vartheta \left[ \begin{array}{cc}
     a  \\
     b 
\end{array} \right](\nu,\tau) = \sum_{l \in \mathbb{Z}} \e^{\im \pi (l +a)^2 \tau} \e^{\im 2\pi (l +a) (\nu +b)},}
and rewrite the zero mode (\ref{torus zero mode 3}) as
\eq{
f^{(p)}_{i_m} (x^{2m-1}, x^{2m}) = \left( \frac{kp}{4 \pi^3 l_{2m}^2}\right)^{1/4} \e^{- \frac{kp}{2} (x^{2m-1})^2} \vartheta \left[ \begin{array}{cc}
     i_m/pq_m  \\
     0 
\end{array} \right]( \frac{pq_m}{2\pi l_{2m}} (x^{2m} - \im x^{2m-1}),pq_m\tau_m).
}
There is the following identity of the theta function \cite{theta},
\als{
\vartheta \left[ \begin{array}{cc}
     r/N_1  \\
     0 
\end{array} \right](N_1 z_1, N_1 \tau) \,
\vartheta \left[ \begin{array}{cc}
     s/N_2  \\
     0 
\end{array} \right](N_2 z_2, N_2 \tau)
= 
\sum_{t=1}^{N_1 + N_2}
\vartheta \left[ \begin{array}{cc}
     \frac{r+s+N_1 t}{N_1 + N_2} 
     \\
     0
\end{array} \right](N_1 z_1 + N_2 z_2, (N_1 + N_2) \tau) &\\
\times
\vartheta \left[ \begin{array}{cc}
     \frac{N_2 r - N_1 s + N_1 N_2 t}{N_1 N_2 (N_1 + N_2)}  \\
     0
\end{array} \right](N_1 N_2 (z_1 - z_2), N_1 N_2 (N_1 + N_2)\tau) &
.}
This implies
\eq{ \label{zero mode products}
f^{(q)}_{j_m} (x) f^{(p)}_{i'_m} (y) = [(p+q)q_m]^{-1/2}\sum_{t=1}^{(p+q)q_m} f^{(p+q)}_{j_m + i'_m + qq_m t} (\tilde{x}) f^{((p+q)pq q_m^2)}_{p q_m j_m - q q_m i'_m + pq q_m^2 t} (\tilde{y}),}
where
\eq{ \tilde{x}^a := \frac{q x^a + p y^a}{p+q}, \quad \tilde{y}^a := \frac{x^a - y^a}{(p+q)q_m}.}

Now, let us calculate the combination $f^{(q)}_{c_m,j_m} (x) f^{(p)}_{i'_m} (y)$, which appears in the integrand of  (\ref{fuzzy torus T_p(f)}).
To do this, we act $a^{(q)\dagger}_m (x)$ on (\ref{zero mode products}) $c_m$ times.
Here, $a^{(q)\dagger}_m (x)$ is the creation operator (\ref{raising and lowering operators}).
From the chain rule of the covariant derivative, we have
\eq{
\label{creation operator identity}
a^{(q)\dagger}_m (x) = \sqrt{\frac{q}{p+q}} a^{(p+q)\dagger}_m (\tilde{x}) + \sqrt{\frac{p}{p+q}} a^{((p+q)pqq_m^2)\dagger}_m (\tilde{y}).}
By using (\ref{Laplacian eigenvector T^2n}) and (\ref{creation operator identity}), we find
\als{ &f^{(q)}_{c_m,j_m} (x) f^{(p)}_{i'_m} (y)\\
&= \sum_{t=1}^{(p+q)q_m} \sum_{c'_m = 0}^{c_m} \sqrt{\frac{c_m !}{(c_m - c'_m)!c'_m!}\frac{q^{c'_m} p^{c_m - c'_m}}{(p+q)^{c_m+1}q_m}}  f^{(p+q)}_{c'_m,j_m + i'_m + qq_m t} (\tilde{x})f^{((p+q)pq q_m^2)}_{c_m - c'_m, p q_m j_m - q q_m i'_m + pq q_m^2 t} ( \tilde{y}).}
By setting $x^a = y^a$, the above equation becomes
\als{
\label{eigenstate products 2}
&f^{(q)}_{c_m,j_m}(x) f^{(p)}_{i'_m} (x) \\
&= \sum_{t=1}^{(p+q)q_m} \sum_{c'_m = 0}^{c_m} \sqrt{\frac{c_m !}{(c_m - c'_m)!c'_m!}\frac{q^{c'_m} p^{c_m - c'_m}}{(p+q)^{c_m+1}q_m}}
f^{(p+q)}_{c'_m,j_m + i'_m + qq_m t} (x) f^{((p+q)pq q_m^2)}_{c_m - c'_m, p q_m j_m - q q_m i'_m + pq q_m^2 t} (0 ).}
By using (\ref{eigenstate products 2}) and (\ref{orthonormality of f_i_m}), we find
\als{&\int_{0}^{2\pi l_{2m-1}} \dif x^{2m-1}  \int_{0}^{2\pi l_{2m}} \dif x^{2m} (f^{(p+q)}_{i_m})^* f^{(q)}_{c_m,j_m} f^{(p)}_{i'_m}\\
&= \sqrt{\frac{p^{c_m}}{(p+q)^{c_m +1} q_m}}\sum_{t=1}^{(p+q)q_m} f^{((p+q)pq q_m^2)}_{c_m, p q_m j_m - q q_m i'_m + pq q_m^2 t} ( 0 ) \, \delta^{(\mathrm{mod} \ (p+q)q_m)}_{i_m, j_m + i'_m + qq_m t}.}
By plugging (\ref{f_c,j}) into the above equation, we finally obtain (\ref{fuzzy torus T_p(f)}).

\end{document}